\documentclass[a4paper,fleqn,usenatbib]{mnras}

\usepackage[T1]{fontenc}
\usepackage{ae,aecompl}

\usepackage{graphicx}	% Including figure files
\usepackage{amsmath}	% Advanced maths commands
\usepackage{amssymb}	% Extra maths symbols
\usepackage{pdflscape}

\newcommand{\tw}{$W_{10\%}$}
\newcommand{\ha}{H$\alpha$}
\newcommand{\kms}{km.s$^{-1}$}

\title[Passive protoplanetary disks in Taurus-Auriga]{A search for passive protoplanetary disks in the Taurus-Auriga star-forming region}

\author[G. Duch\^ene et al.]{
Gaspard Duch\^ene,$^{1,2}$\thanks{E-mail: gduchene@berkeley.edu}
Adam Becker,$^{1}$
Yizhe Yang,$^{3}$
Herv\'e Bouy,$^{4,5}$
Robert J. De Rosa,$^{1}$
\newauthor Jennifer Patience$^{6}$, Julien H. Girard$^{7,2}$\\
$^{1}$Astronomy Department, University of California,
    Berkeley, CA 94720, USA\\
$^{2}$Universit\'e Grenoble Alpes, CNRS, IPAG, Grenoble, 38000, France\\
$^{3}$Physics Department, University of California,
    Berkeley, CA 94720, USA\\
$^{4}$Laboratoire d'Astrophysique de Bordeaux, Univ. Bordeaux, CNRS, B18N, all\'ee Geoffroy Saint-Hilaire, 33615 Pessac, France\\
$^{5}$Centro de Astrobiolog\'{\i}a, Depto de Astrof\'{\i}sica, INTA-CSIC, PO BOX 78, 28691, ESAC Campus, E-208691 Villanueva de la Ca\~{n}ada,\\ Madrid, Spain\\
$^{6}$School of Earth and Space Exploration, Arizona State University, P.O. Box 871404, Tempe, AZ 85287, USA\\
$^{7}$ European Southern Observatory, Casilla 19001, Santiago 19, Chile}

\date{Accepted XXX. Received YYY; in original form ZZZ}

\pubyear{2016}

\begin{document}
\label{firstpage}
\pagerange{\pageref{firstpage}--\pageref{lastpage}}
\maketitle

\begin{abstract}
We conducted a 12-month monitoring campaign of 33 T Tauri stars (TTS) in Taurus. Our goal was to monitor objects that possess a disk but have a weak \ha\ line, a common accretion tracer for young stars, to determine whether they host a passive circumstellar disk. We used medium-resolution optical spectroscopy to assess the objects' accretion status and to measure the \ha\ line. We found no convincing example of passive disks; only transition disk and debris disk systems in our sample are non-accreting. Among accretors, we find no example of flickering accretion, leading to an upper limit of 2.2\% on the duty cycle of accretion gaps assuming that all accreting TTS experience such events. Combining literature results with our observations, we find that the reliability of traditional \ha-based criteria to test for accretion is high but imperfect, particularly for low-mass TTS. We find a significant correlation between stellar mass and the full width at 10 per cent of the peak (\tw) of the \ha\ line that does not seem to be related to variations in free-fall velocity. Finally, our data reveal a positive correlation between the \ha\ equivalent width and its \tw, indicative of a systematic modulation in the line profile whereby the high-velocity wings of the line are proportionally more enhanced than its core when the line luminosity increases. We argue that this supports the hypothesis that the mass accretion rate on the central star is correlated with the \ha\ \tw\ through a common physical mechanism.
\end{abstract}

\begin{keywords}
stars: variables: T Tauri, Herbig Ae/Be -- circumstellar matter
\end{keywords}

%%%%%%%%%%%%%%%%%%%%%%%%%%%%%%%%%%%%%%%%%%%%%%%%%%

%%%%%%%%%%%%%%%%% BODY OF PAPER %%%%%%%%%%%%%%%%%%

\section{Introduction}

T\,Tauri stars (TTS) are pre-main sequence stars with masses $M \lesssim 1.5\,M_\odot$. Many of these objects host circumstellar disks that can be traced by the thermal emission of the dust they contain \citep{haisch01, ribas14}. Sensitive infrared surveys with the {\it Spitzer} and {\it Herschel} observatories have identified thousands of TTS with fully sampled spectral energy distributions (SEDs) in nearby star-forming regions and enabled statistical studies of the evolution of protoplanetary disks \citep[][and references therein]{williams11, alexander14}. While most TTS present a full-fledged excess extending from the near-infrared to the millimeter regime, these surveys have also revealed a number of objects whose excess is significantly subdued, sometimes even completely absent, in the near- to mid-infrared regime. Transition disks, in which the inner regions of the disk are almost entirely devoid of dust (and gas), suggest that the opening of an inner hole is an important step in the disk dissipation process although the debate about the physical mechanism(s) responsible for this hole opening is still ongoing \citep{espaillat14}.

Dissipative forces in the gaseous component of protoplanetary disks are responsible for angular momentum transport in these disks, which ultimately leads to accretion on the central stars \citep{bertout89, hartmann16}. This accretion process is believed to be magnetically mediated and, as a side effect, is also responsible for the launching of collimated jets \citep{bouvier07}. TTS displaying evidence for accretion and mass loss are referred to as ``classical'' TTS (CTTS). Equally young pre-main sequence stars with no evidence for ongoing accretion are  dubbed ``weak-lined'' TTS (WTTS). For magneto-accretion to proceed, the circumstellar disk must extend all the way in to the co-rotation radius, located just a few stellar radii away from the star. In this picture, if an inner hole is carved out as part of disk evolution, one expects accretion to stop. Indeed, in an inside-out disk clearing picture, one expects accretion to stop before a significant signature builds up in the system's SED, specifically a deficit of near-infrared excess. In other words, while the presence of both a protoplanetary disk and ongoing accretion have been assumed to systematically come hand-in-hand, it is possible that some objects may show one but not the other. Indeed, the  correlation between active accretion and the presence of a full-fledged disk has proven to be less automatic than initially expected. A number of significant infrared excess have been discovered among WTTS \citep{padgett06, cieza13, fang13} and some transitional disks have been found to accrete at a rate comparable to that of normal CTTS \citep[e.g.,][]{najita07}. The former category is particularly intriguing as these ``passive'' (i.e., non-accreting) circumstellar disks could represent a key initial stage of disk dissipation \citep[e.g.,][]{mccabe06}.

The energy deposited by accretion on the star is radiated away via a combination of hot continuum and line emission. As a consequence of the characteristic 10$^4$\,K temperature of accretion hotspot, most of the continuum is emitted at ultraviolet wavelengths and observations in this regime are the most sensitive and best quantitative tracers of the mass accretion rate \citep{gullbring98, herczeg08, rigliaco11, ingleby13}. However, the intrinsically red colors of TTS and the common presence of significant line-of-sight extinction often preclude using this diagnostic of accretion. Instead, line emission is the most commonly used tracer of accretion in TTS. While many lines throughout the optical and near-infrared wavelength range correlate tightly with UV-determined accretion rates \citep{antoniucci11, rigliaco12}, the \ha\ line remains the most common tracer of accretion by virtue of it being the strongest emission line, despite significant ontribution from chromospheric activity and stellar winds and outflows. In particular, TTS have long been characterized as accreting if their \ha\ equivalent width (EW) exceeds a spectral type-dependent threshold set to distinguish them from chromospherically active stars \citep[][and references therein]{barrado03}. However, it was later suggested that the intrinsic linewidth, specifically \tw, is a more accurate discriminator between accreting and non-accreting TTS \citep{white03, natta04} as chromospheric activity is generally characterized by much lower velocities than organized accretion flows. It must be pointed out, however, that different studies yield somewhat different accretion thresholds \citep{cieza13, fang13}. Nonetheless, the observed correlation between \tw\ and the mass accretion rate has led to the notion that measuring the line width of \ha\ can provide a quantitative estimate of the accretion rate \citep{natta04}. Considering the line-of-sight velocities involved, medium- to high-resolution spectroscopy is necessary to measure \tw, so that this quantity is generally not probed when performing reconnaissance surveys of TTS, leaving the EW as the common, albeit imperfect, criterion used to determine the accretion status of TTS.

Accretion on young stellar objects is a highly variable phenomenon. Indeed, photometric variability was one of the original criteria to identify such objects \citep{joy45, herbig60}. Leaving aside the most extreme FU\,Ori-like outbursts \citep{audard14}, TTS are known to display variability in their accretion rate on the order of 0.5\,dex on timescales ranging from hours to years \citep{alencar02, nguyen09, fang13, costigan14, venuti14}. In this context, it is plausible that the apparent conundrum posed by WTTS with full fledged disks could be a consequence of some extreme cases of variability. If the accretion rate varies so much as to effectively flicker on and off, single epoch probes of accretion may not yield the full picture for any particular object. However, TTS for which the accretion flow is known to temporarily pause altogether have proved remarkably elusive \citep[e.g.,][]{costigan12}, with only one relatively old ($\sim8$\,Myr) TTS being flagged as undergoing episodic accretion to date \citep{murphy11}. 

An alternative interpretation of the apparently passive disks is that past spectroscopic classification was inaccurate. It is possible, for instance, that self-absorption in the \ha\ line profile, a common occurrence among CTTS, remains undetected in low-resolution spectra, resulting in a low EW for the emission line \citep{fang13}. In such cases, the level of accretion can be more readily assessed by measuring the \ha\ \tw\ if its correlation with accretion rate is rooted in a physical mechanism. This could be particularly valuable in the case of weakly accreting TTS, for which the UV excess diagnostic is not easy to probe, especially for M-type objects \citep{venuti14}.

Here we present the results of a 12-month medium-resolution spectroscopic monitoring campaign aimed at identifying  permanently passive disks and/or disks experiencing flickering accretion in a sample of 31 disk-bearing TTS in the Taurus star-forming region. The targets were selected to represent as broad a diversity of accretion states as possible, from apparently non-accreting WTTS to strongly accreting CTTS. This paper is organized as follows: Sect.\,\ref{sec:method} discusses the choices made in the practical implementation of this description, and Sect.\,\ref{sec:obs} presents the observations and data reduction details. The main results of this survey and their analysis are presented in Sect.\,\ref{sec:results}. Finally, we discuss the lack of confirmed passive disks, the general variability trends observed in this study, and the pertinence of the \ha\ in assessing the accretion status of TTS in Sect.\,\ref{sec:discuss}. Sect.\,\ref{sec:conclu} summarizes our main results.

%________________________________________________________________

\section{Survey methodology}
\label{sec:method}

%_ _ _ _ _ _ _ _ _ _ _ _ _ _ _ _ _ _ _ _ _ _ _ _ _ _ _ _ _ _ _ _ _ _ _ _ _ _ _ _ _ 

\subsection{Set-up of the campaign}
\label{subsec:setup}

% . . . . . . . . . . . . . . . . . . . . . . . . . . . . . . . . . . . . . . . . . . . . . . . . . . . . . 
\subsubsection{Instrumental configuration}

While observations dedicated to accurately measuring the \ha\ line profile and \tw\ use high-resolution spectroscopy ($R \gtrsim 25,000$), such settings require relatively long integration times that are not amenable to extensive monitoring campaigns of large samples. To circumvent this issue, and obtain as many epochs as possible, we have instead adopted a strategy based on a moderate spectral resolution. To estimate the spectral resolution necessary to achieve our goal, we use a toy model in which the \ha\ line is approximated by a Gaussian profile, which is reasonable for objects with relatively narrow emission lines. In this situation, the observed \tw\ of the line and its full width at half maximum (FWHM) are connected via the relationship $W_{10\%} \approx 1.82$\,FWHM. Thus, a \tw\ of 200\,\kms\ \citep[the lowest proposed threshold demarcating accreting from non-accreting TTS,][]{natta04} corresponds to a FWHM of 110\,\kms. Such a line can be spectrally resolved if $R \gtrsim 2,700$ (achieved during the majority of our observations), assuming high signal-to-noise ratio. This toy model, which neglects the effects of instrumental line broadening and limited spectral sampling, served as a coarse guideline for selecting the instrumental set-ups used in the survey. In Sect.\,\ref{subsec:simul}, we present a quantitative analysis of the accuracy of the \tw\ measurements using actual data.

While the primary focus of the survey is the \ha\ line (see below), other accretion-related emission lines are present in optical spectra of TTS \citep[e.g.,][]{herczeg08, fang09, rigliaco12}, providing additional elements of analysis in cases where \ha\ measurements are ambiguous \citep{jayawardhana06}. In particular, we make use of the 5876 and 6678\,\AA\ He\,I emission lines. In addition, when a wide spectral coverage is available, we also include the 8498--8662\,\AA\ Ca triplet, also directly connected to accretion in TTS. Finally, we also consider the 6300\,\AA\ [O\,I] forbidden emission line, one of strongest tracers of mass loss in young low-mass stars that indirectly reveals underlying accretion on the central object \citep{hartigan95}. We note, however, that the detection of the [O\,I] line can be ambiguous at the medium resolution adopted here, as atmospheric glow also produces line emission. We thus put less stock in that line than on the He\,I and Ca lines.

% . . . . . . . . . . . . . . . . . . . . . . . . . . . . . . . . . . . . . . . . . . . . . . . . . . . . . 
\subsubsection{Temporal sampling}

Since accretion is a highly variable process, it is necessary to obtain a synoptic dataset covering a broad range of timescales to accurately assess the status of any given TTS \citep{nguyen09, costigan14}. To capture this behavior, we have devised our monitoring campaign to probe, roughly uniformly, a broad range of timescales, from 1\,day up to $\sim$1\,yr. The latter upper limit was selected to roughly match the dynamical (orbital) timescale at a distance of 1\,au from the central source. Persistent accretion (or absence thereof) over such a timescale indicates that, even though the accretion flow may experience short-term fluctuations, the overall inward flow of material within that region remains stable (or absent) over dynamically long timescales.

%_ _ _ _ _ _ _ _ _ _ _ _ _ _ _ _ _ _ _ _ _ _ _ _ _ _ _ _ _ _ _ _ _ _ _ _ _ _ _ _ _ 

\subsection{Sample selection}
\label{subsec:sample}

\begin{figure*}
\includegraphics[width=0.49\textwidth]{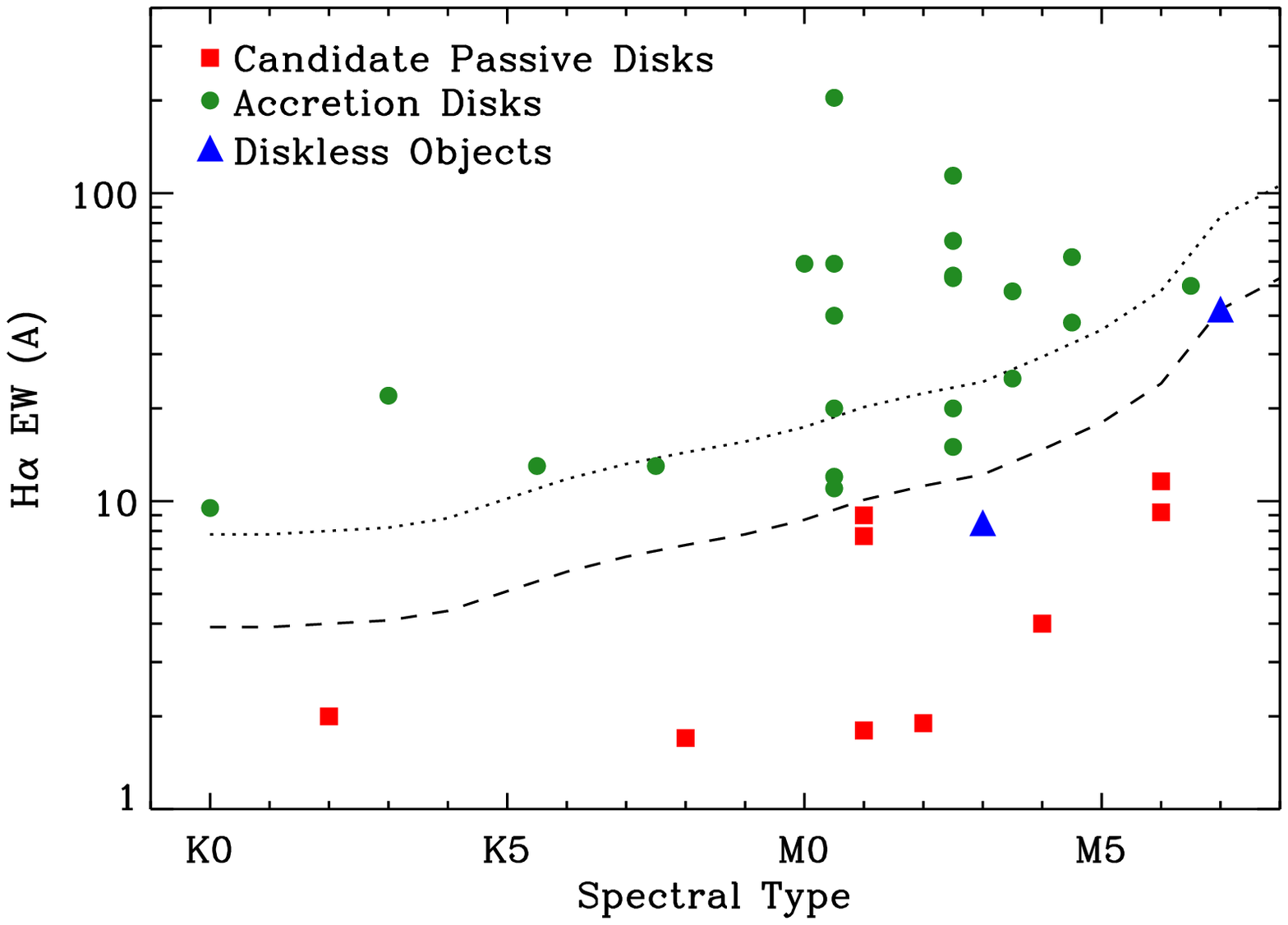}
\includegraphics[width=0.49\textwidth]{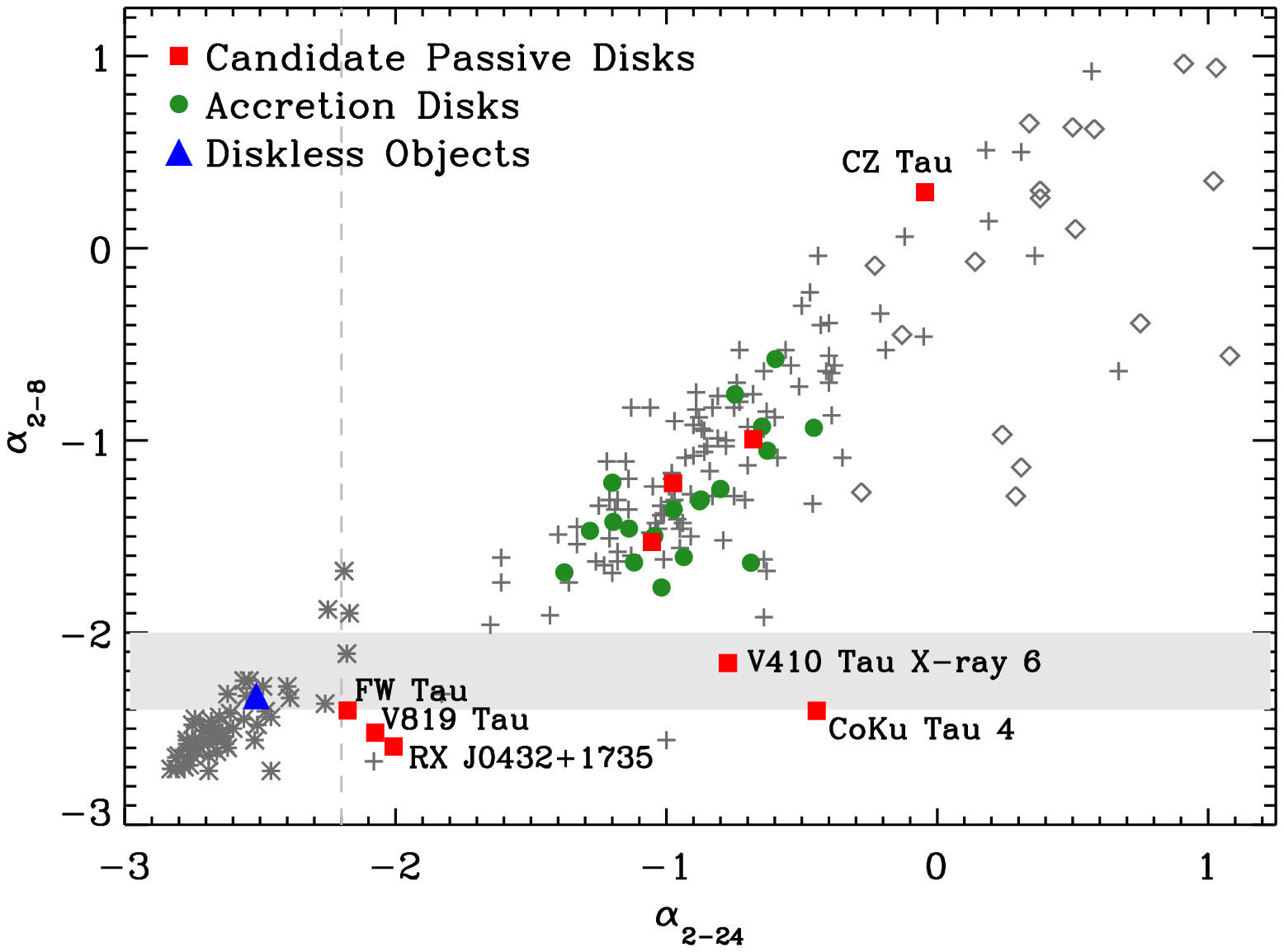}
\caption{
{\it Left:} \ha\ EW from the literature as a function of spectral type for the targets in our sample. The dashed and dotted line represent the canonical threshold for separating accreting from non-accreting T\,Tauri stars \citep{barrado03} and twice that value, respectively. Objects lying between those two lines are refered to as ``borderline'' objects since they lie close to the threshold value, while those above the dotted line make up our control sample of strongly accreting objects. Red squares and green circles indicate candidate passive disks and accreting objects, respectively. The two blue triangles are disk-less objects that we monitored to serve as benchmark for non-accreting low-mass T\,Tauri stars. {\it Right:} Power law indices to the mid-infrared part of the SED of members of the Taurus star-forming region, from \citet{luhman10} : asterisks, plus signs and diamonds represent known Class III, II and I, respectively. The samples targeted in this study are indicated by filled green circles (``borderline'' and control sample targets), red squares (candidate passive disks) and blue triangles (disk-less targets). The gray horizontal band and vertical line mark the threshold for assessing the presence of significant infrared excess, as defined by \citet{luhman10}. Five targets have incomplete {\it Spitzer} photometry and cannot be placed in this diagram. Their inclusion in our survey is discussed in Sect.\,\ref{subsec:sample}. 
}\label{fig:sample}
\end{figure*}

This survey is focused on the Taurus-Auriga star-forming region, arguably the best-studied population of TTS among nearby molecular clouds. Understanding the frequency at which passive disks and objects with flickering accretion are present in that region will provide valuable insight for studies of more distant star-forming regions, where most often only single-epoch, modest spectral resolution studies are available. 

Our sample selection started with the identification of objects with a confirmed circumstellar disk, for which we performed a thorough literature search of \ha\ EW measurements. Based on this, we created the following 3 subsamples:

\begin{itemize}
\item[--] {\it candidate passive disk objects} (9 targets), whose published \ha\ EW measurements fall below the standard accretion threshold defined by \cite{barrado03};
\item[--] {\it borderline objects} (7 targets), whose \ha\ EW is within 1 and 2 times this threshold;
\item[--] a {\it control sample} (15 targets) of accreting TTS with strong \ha\ EW.
\end{itemize}

In addition to these 31 targets with a confirmed disk, we also observed 2 disk-less objects (UX\,Tau\,C and MHO\,4) with a late spectral type in order to test the fiability of our \ha\ measurements. Indeed, in the case of weak emission lines superimposed on a continuum dominated by deep photospheric features, larger uncertainties are associated with these measurements due to uncertainties in estimating the level of continuum emission at our intermediate spectral resolution. 

Each of the subsamples defined above addresses a distinct question. The candidate passive disks are the most intriguing and we wish to confirm the prima facie interpretation of their published properties as being non-accreting circumstellar disk. If confirmed, they would be in a rather unique state. Alternatively, it may be that newer spectroscopic measurements, particularly considering the \tw\ quantity, leads to a re-evaluation of the objects. The borderline objects were selected to test the hypothesis that accretion may be flickering in objects with a relatively low accretion rate (although we note that a moderate \ha\ EW does not necessarily imply a low accretion rate). Finally, the control sample was observed to understand the typical variability of ``normal'' TTS over the timescales and with the observational probes used in this study.

As shown in Figure\,\ref{fig:sample}, most of our targets have near- and mid-infrared colors typical of full-fledged TTS disks. A few targets with little near-infrared ($[K-8]$) excess but modest-to-strong mid-infrared ($[K-24]$) excess, generally classified as transition disks, are included in this survey. For a handful of targets, even though the {\it Spitzer} photometry is incomplete, \cite{luhman10} were able to classify the objects as ``class II'', with the exceptions of FW\,Tau and RX\,J0432+1735. For these two objects, significant mid-infrared, far-infrared excess and/or millimeter emission from circumstellar dust has been confirmed \citep{andrews05, wahhaj10, cieza13, howard13}. On the other hand, CZ\,Tau appears to have extremely red colors, more typical of an embedded Class\,I source (this source is further discussed in Sect.\,\ref{subsec:discuss_passive}). 

We further note that the distribution of spectral types in our sample, from K0 to M6.5, is characteristic of the entire Taurus population \citep[e.g.,][]{luhman06}. Similarly, the ranges of visible brightness are similar for all three subsamples. Finally, as is typical for any sample of pre-main sequence stars in Taurus, a significant fraction of our targets (12/33) are members of multiple systems, with separations ranging from 75 to 2000\,au. Systems with projected separation below $\approx2$\arcsec\ are liable to source confusion in our seeing-limited observations, and care must be taken when interpreting the system-integrated spectra we obtain here (see Sect.\ref{sec:results}). 

As discussed above, the \ha\ line is not guaranteed to be a perfect tracer of accretion, so the W/C TTS classification can only be considered a first attempt at assessing the accretion status of our targets. Fortunately, the rich literature on the Taurus-Auriga star-forming region provides us with more discriminating observations, such as the presence of UV excess and/or veiling of photospheric lines across the visible range. We explored the results of UV excess searches \citep{muzerolle98, white01, valenti03, ingleby13}, veiling in the blue optical \citep{valenti93}, and veiling in the red optical \citep{white03, herczeg08, herczeg14} for all of our targets. If a target was found to have evidence for accretion in any of these observations, it is classified as an a priori accretor. In the absence of observations in the UV or blue optical, the veiling in the red optical is the only available element. However, since no red veiling is sometimes found in UV-confirmed accretors, the classification based on red veiling alone is considered uncertain in case of a non-detection. Finally, we note that MHO\,5 is classified as a "possible accretor" by \cite{herczeg08}, i.e., its status remains uncertain.

The main properties of all sources monitored in this program are presented in Table\,\ref{tab:sample}.

%\begin{landscape}
\begin{table*}
\caption{Observed sample. Spectral types and SED class are adopted from \citet{herczeg14} unless otherwise stated. The literature TTS classification is based on the strength of the \ha\ line relative to the spectral type-dependent accretion threshold defined by \citet{barrado03}. The presence of UV excess and/or optical veiling is taken from the literature (see Sect.\,\ref{subsec:sample}). The multiplicity flag indicates single (S), binary (B), triple (T) and quadruple (Q) systems; only companions out to projected separations up to 2000\,AU are considered. $N_{spec}$ is the number of spectra obtained in this study, separated in time by a maximum of $\Delta t_{max}$. References for \ha\ measurements: (1) \citet{hbc88}; (2) \citet{white01}; (3) \citet{wahhaj10}; (4) \citet{strom94}; (5) \citet{beckwith90}; (6) \citet{briceno98}; (7) \citet{hartigan94}; (8) \citet{nguyen09}. Individual notes: (a) Far-infrared and mm excess \citep{andrews05, howard13}; (b) \citet{wichmann96}; (c) 24$\mu$m and 70$\mu$m excess \citep{wahhaj10, cieza13}; (d) \citet{luhman10}.}
\label{tab:sample}
\begin{tabular}{lccccccccccc}
\hline
Object & Sp.T. & $R$ & EW(\ha) & H$\alpha$ Ref. & \multicolumn{3}{c}{Litt. Classification} & Accretor & Mult. & $N_{spec}$ & $\Delta t_{max}$ \\
 & & [mag] & [\AA] & & TTS & UV/veiling & SED & in this survey & & & [days] \\
\hline
\multicolumn{10}{c}{Candidate passive disks}\\
\hline
CoKu Tau 4 & M1 & 14.48 & -1.8 & 1 & W & N? & II & N & B & 11 & 372 \\  % M1.1
CZ Tau & M4 & 14.77 & -4.0 & 1 & W & N & II & N? & B & 11 & 373 \\  % M4.2
FW Tau & M6 & 14.80 & -11.6 & 2 & W & N & II/III$^{\mathrm a}$ & N & T & 11 & 372 \\  % M5.8
HQ Tau & K2 & 11.18 & -2 & 8 & W & N? & II & Y & S & 11 & 372 \\  % K2 
IQ Tau & M1 & 12.28 & -7.7 & 1 & W & Y & II & Y & S & 8 & 60 \\  % M1.1
RX J0432.8+1735& M2$^{\mathrm b}$ & \dots & -1.9 & 3 & W & \dots & II$^{\mathrm c}$ & N & S & 11 & 372 \\
V410 Tau X-ray 6 & M6 & 15.08 & -9.2 & 4 & W & \dots & II & N? & S & 10 & 372 \\  % M5.9 
V819 Tau & K8 & 12.24 & -1.7 & 1 & W & Y & II & N & S & 11 & 372 \\  % K8 
V836 Tau & M1 & 12.17 & -9 & 1 & W & Y & II & Y & S & 11 & 373 \\ % a.k.a. P5 in Mundt et al. (1983) - variable Halpha, EW = 5 .. 15 A, SpT = K7-M0  % % M0.8 
\hline
\multicolumn{10}{c}{Borderline objects}\\
\hline
CX Tau & M2.5 & 12.65 & -20 & 1 & C & Y & II & Y & S & 8 & 60 \\  % M2.5 
DN Tau & M0.5 & 11.49 & -12 & 1 & C & Y & II & Y & S & 9 & 370 \\  % M0.3 
FN Tau & M3.5 & 14.68 & -25 & 1 & C & N & II & Y? & S & 6 & 361 \\  % M3.5 
GH Tau & M2.5 & 11.77 &  -15 & 2 & C & Y & II & Y & B &5 & 57 \\  % M2.3 
IP Tau & M0.5 & 12.46 & -11 & 1 & C & Y & II & Y & S & 8 & 60 \\  % M0.6 
MHO 5 & M6.5 & 16.23 & -50 & 6 & C & Y? & II & Y & S & 10 & 373 \\  % M6.5
V807 Tau & K7.5 & 10.07 & -13 & 1 & C & Y & II & Y & T & 8 & 60 \\  % K7.5 
\hline
\multicolumn{10}{c}{Control sample - Disk}\\
\hline
BP Tau & M0.5 & 11.62 & -40 & 4 & C & Y & II & Y & S & 8 & 60 \\  % M0.5 
CY Tau & M2.5 & 12.35 & -70 & 1 & C & Y & II & Y & S & 8 & 60 \\  % M2.3 
DE Tau & M2.5 & 11.69 & -54 & 4 & C & Y & II & Y & S & 8 & 60 \\  % M2.3 
DH Tau & M2.5 & 13.59 & -53 & 1 & C & Y & II & Y & B &7 & 369 \\  % M2.3 
DS Tau & M0.5 & 11.56 & -59 & 1 & C & Y & II & Y & S & 8 & 60 \\  % M0.4 
FM Tau & M4.5 & 13.64 & -62 & 7 & C & Y & II & Y & S & 9 & 370 \\  % M4.5 
FP Tau & M2.5 & 12.72 & -38 & 1 & C & Y & II & Y & S & 8 & 60 \\  % M2.6 
FQ Tau & M4.5 & 13.68 & -114 & 4 & C & Y & II & Y & B &5 & 57 \\  % M4.3 
FY Tau & M0 & 14.17 & -59 & 1 & C & Y & II & Y & S & 8 & 60 \\  % M0.1 
FZ Tau & M0.5 & 13.83 & -204 & 1 & C & Y & II & Y & S & 4 & 57 \\  % M0.5 
GI Tau & M0.5 & 12.15 & -20 & 7 & C & Y & II & Y & T & 8 & 60 \\  % M0.4 
GK Tau & K3 & 11.58 & -22 & 7 & C & Y & II & Y & T &8 & 60 \\  % K3 
LkCa 15 & K5.5 & 11.61 & -13 & 1 & C & Y & II & Y & S & 8 & 60 \\  % K5.5 
UX Tau A & K0 & 10.48 & -9.5 & 2 & C & Y & II & Y & Q & 9 & 371 \\  % K0 
V710 Tau & M3.5 & 12.50 & -48 & 1 & C & N? & II & Y & B & 8 & 370 \\  % M3.3 
\hline
\multicolumn{10}{c}{Control sample - Disk-less}\\
\hline
MHO 4 & M7$^{\mathrm d}$ & 16.36 & -42 & 6 & W & N? & III & Y? & S & 11 & 373 \\
UX Tau C & M3 & 15.11 & -8.5 & 2 & W & N? & III & N & Q & 7 & 60 \\  % M2.8 
\hline
\end{tabular}
\end{table*}
%\end{landscape}

%________________________________________________________________

\section{Observations and data reduction}
\label{sec:obs}

\begin{table*}
\caption{Instrumental setup. $W_{\mathrm{10\%}}^{\mathrm{inst}}$ indicates the measured width at 10 per cent of the peak of isolated emission lines around the wavelength of the \ha\ line from the spectra of calibration (arc) lamps.}\label{tab:setup}
\begin{tabular}{ccccccc}
\hline
Telescope\vspace*{-0.1cm} & Instrum. + Grating & Bandpass [\AA] & Sampling [\AA] & $R$ & $W_{\mathrm{10\%}}^{\mathrm{inst}}$ [\kms] & Obs. Date \\
%&  & [\AA] & [\AA] & & [\kms] & \\
\hline
CAHA 3.5m & TWIN + T06 & 5870--7030 & 0.55 & 4860 & 135$\pm$5 & 2009/10/27 \\
CAHA 3.5m & TWIN + T07 & 5870--9100 & 1.64 & 2020 & 335$\pm$10 & 2009/11/23, 2009/12/10 \\
WHT & ISIS + R600R & 5720--7800 & 0.49 & 4560 & 125$\pm$5 & 2009/10/30--31, 2009/12/26 \\
WHT & ISIS + R158R & 4780--9700 & 1.82 & 400 & 900$\pm$30 & 2009/11/07 \\
SPM 2.1m & Boller\&Chivens & 5400-7600 & 2.07 & 1780 & 360$\pm$5 & 2009/12/02--03 \\
Lick 3m & KAST + \#4 & 5800--7200 & 1.18 & 2770 & 260$\pm$10 & 2010/11/11--14 \\
\hline
\end{tabular}
\end{table*}

The observations were gathered at the Calar Alto (CAHA) 3.5m telescope, the 4.2m William Herschel Telescope (WHT), the San Pedro Martir (SPM) 2.1m telescope and the Lick Observatory Shane 3m telescope, using their respective optical spectrographs. The gratings used for the observations, and the resulting spectral resolution (as well as the \tw\ measured on isolated lines in the spectra of arc lamps), are summarized in Table\,\ref{tab:setup}. The spectral resolution of the observations obtained with the WHT/ISIS R185R grating is too low to measure any intrinsic line width and we only measure the EW of emission lines. Slit widths of 1\arcsec\ to 2\arcsec\ were used, depending primarily on seeing conditions. Apart for the cases of close binaries, the slit was oriented along the paralactic angle to minimize the effects of differential atmospheric refraction. 

Data were obtained during nine separate nights in Fall 2009, supplemented by a 4-night run in Fall 2010 during which the most interesting objects were observed to achieve a 1\,yr total time baseline. The UX\,Tau A--C pair (2\farcs7 separation) could only be resolved during the nights with the best seeing, resulting in poorer sampling for the faint secondary. Overall, our survey covers timescales ranging from 1 to $\approx 60$\,d for most targets, almost uniformly, in addition to the 1\,yr longer baseline for the objects observed in Fall 2010. In some cases, multiple spectra of the same target were obtained consecutively to improve on-sky efficiency. However, because these are always for bright targets with high signal-to-noise, we do not use them to probe variability of timescales $\lesssim$5\,min; we only used these spectra to reject cosmic-ray hits and uncorrected bad pixels.

The raw data products from all instruments used in this survey are very similar in their structure. We thus used the same data reduction pipeline to all datasets, with only minor adjustments. First, a bias frame was subtracted from the raw spectral frames before a flat field correction (based on the spectrum of a broad spectral lamp in the dome or within the calibration unit of the instrument) was applied. Second, the raw one-dimensional spectrum of the target was extracted from the two-dimensional frame using a fixed-width window centered on the position of the star at each pixel along the dispersion axis. An estimate of the sky background emission, determined using the median of adjacent pixels along the slit direction, was then subtracted, taking advantage of the fact that the spatial and spectral axes are nearly perpendicular for all spectrographs used in this study. The wavelength calibration was established using the spectra of arc lamps. In the case of the KAST spectrograph, both the flat-field correction and the wavelength calibration had to be conducted for every individual target to compensate for the effects of flexures, which are particularly large in the red arm of this instrument. 

The spectra were then corrected for airmass using atmospheric absorption profiles appropriate for each observatory. This step is not aimed at fully suppressing all telluric absorption features but rather to remove the overall color trend induced by the atmosphere, by bringing all spectra to a uniform airmass (of unity). Finally, atmospheric absorption features were estimated from the ratio of the observed to the absolute spectra of several early-type spectroscopic standards. While such an approach results, in principle, in an absolute spectrophotometric calibration, our observing strategy (e.g., fixed slit width throughout the night despite sometimes varying seeing conditions, slit orientation not fixed to the parallactic angle) as well as the occasional presence of thin clouds, precluded such a calibration. Our program, which is focused on the width and EW of emission lines but not on their absolute luminosity, nor on the precise continuum shape of each spectrum, is insensitive to the approximative precision of this process. The final spectra usually achieve high signal-to-noise ratio (SNR) in the vicinity of the \ha\ line, with 96\%, 90\% and 80\% or our spectra exceeding SNR=50, 75 and 100 per resolution element, respectively (see Table\,\ref{tab:res}).

\begin{figure*}
\includegraphics[width=\columnwidth]{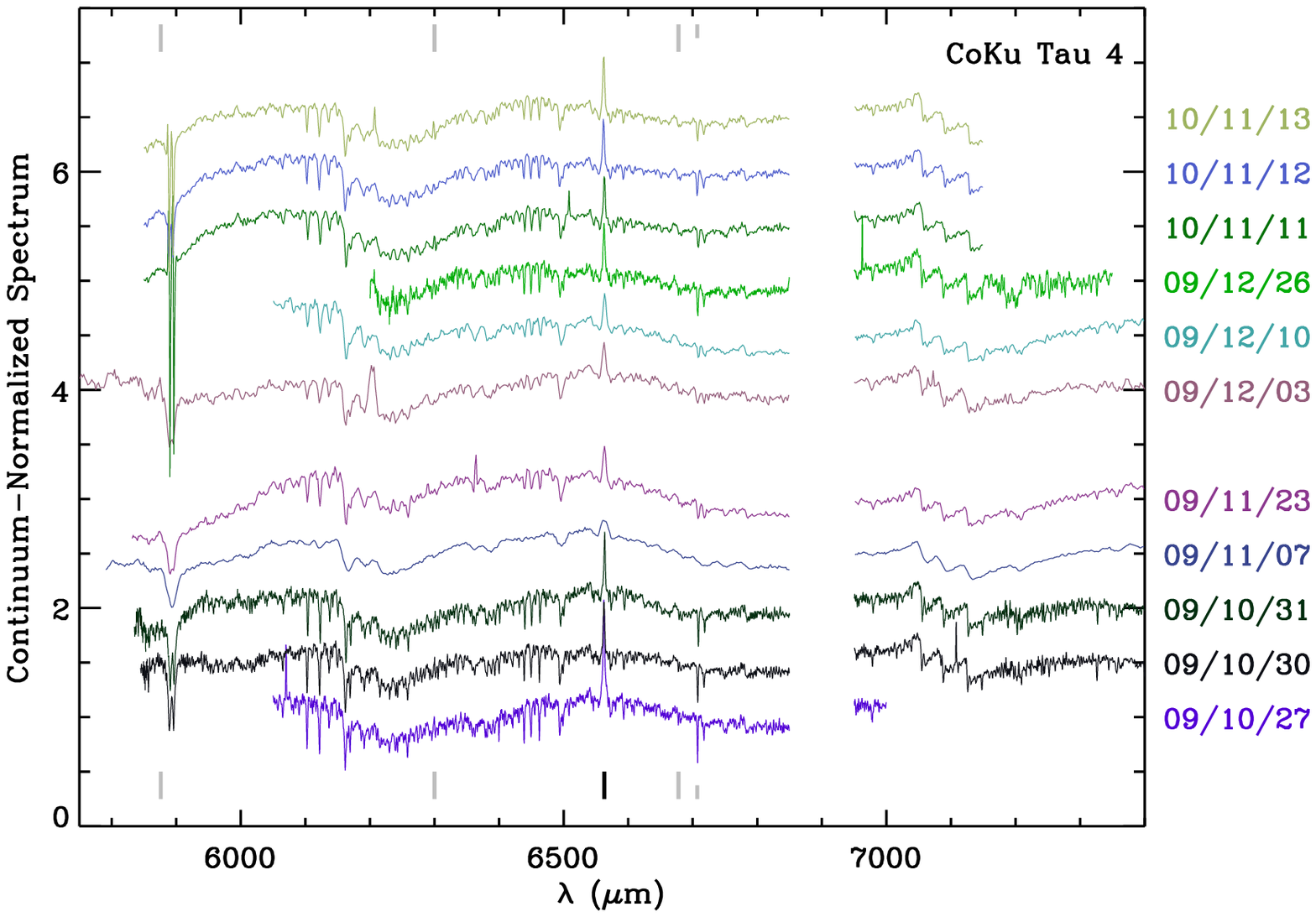}
\includegraphics[width=\columnwidth]{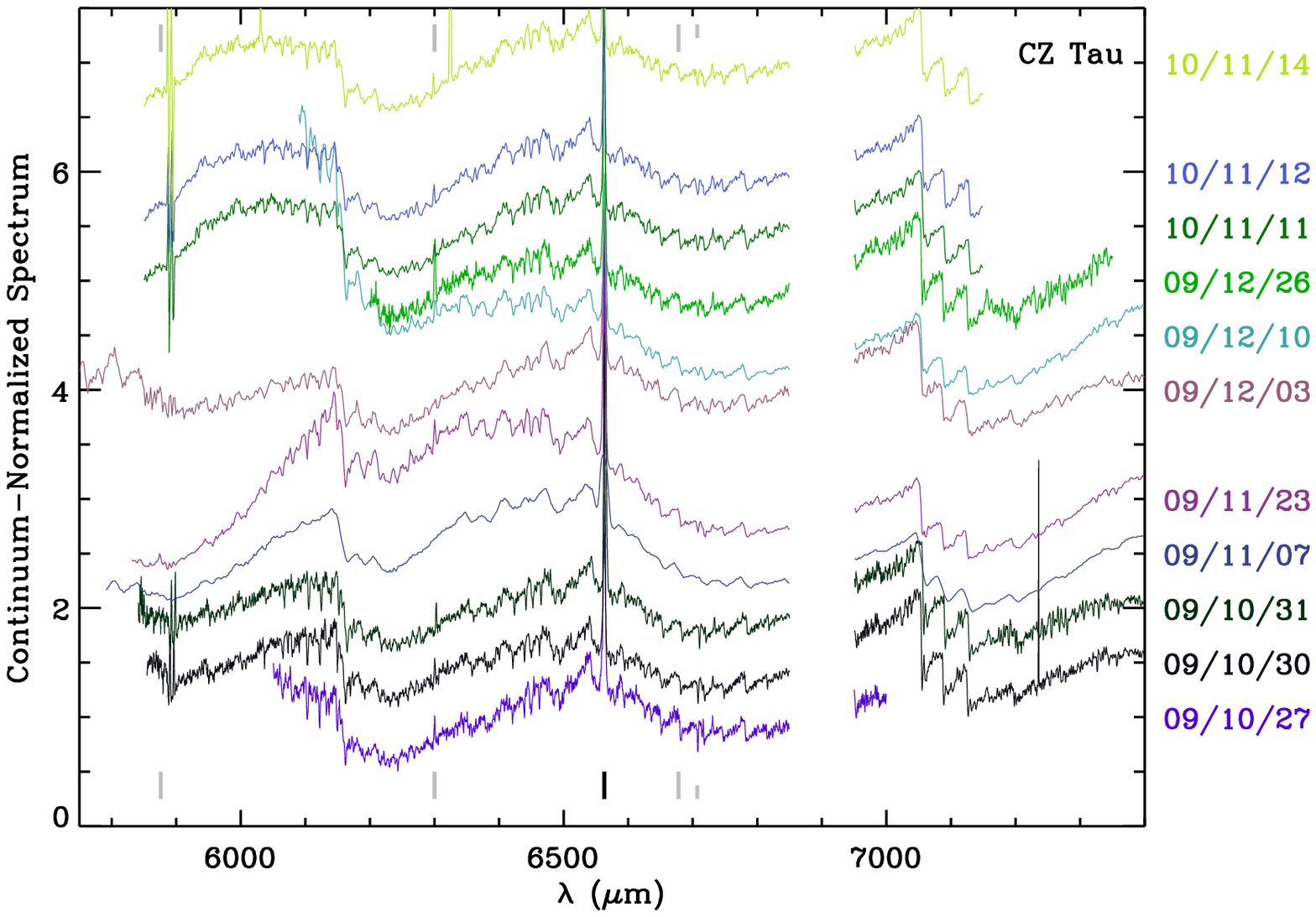}\\
\includegraphics[width=\columnwidth]{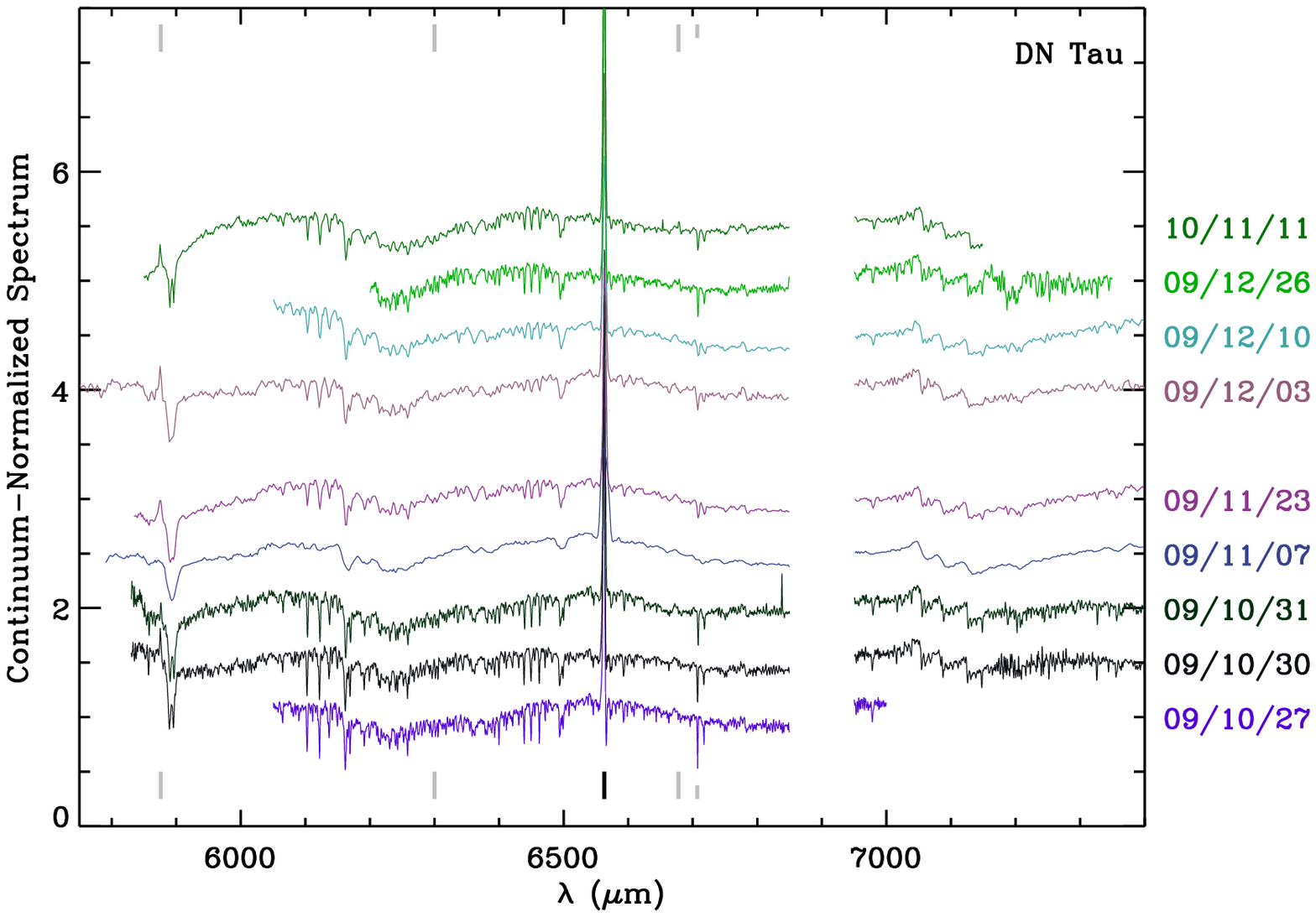}
\includegraphics[width=\columnwidth]{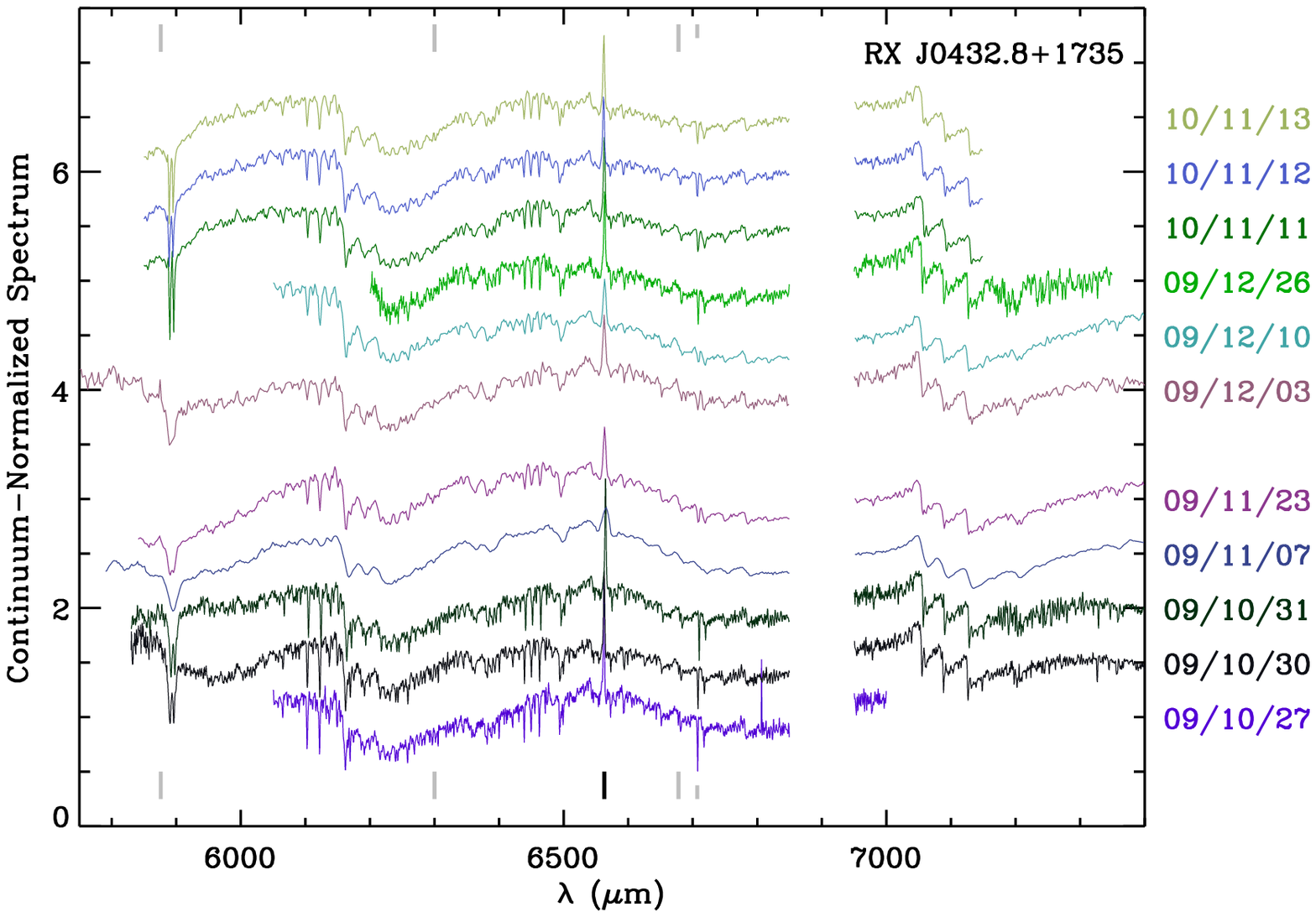}
\caption{All spectra for four of the targets in this survey, each normalized by a low order polynomial function. Vertical tick marks indicate the \ha\ (long black tick), the 6300\,\AA\ [O\,I], 5876\,\AA\ and 6678\,\AA\, He\,I emission lines (long gray ticks) and the 6707\,\AA\ Li photospheric absorption line (short gray tick). The gap around 6900\,\AA\ corresponds to a region of poor atmospheric transmission in which data quality is severely affected. In a few cases, the polynomial fit was imperfect, resulting in incorrect continuum slope (e.g., CZ\,Tau on 2009 Nov 23).}\label{fig:all_specs}
\end{figure*}

%________________________________________________________________

\section{Results}
\label{sec:results}

%_ _ _ _ _ _ _ _ _ _ _ _ _ _ _ _ _ _ _ _ _ _ _ _ _ _ _ _ _ _ _ _ _ _ _ _ _ _ _ _ _ 

\subsection{Raw emission line measurements}
\label{subsec:measur}

The primary spectral feature studied here is the \ha\ emission line, which is detected in all spectra of all sources. To derive its EW and \tw, the continuum was interpolated linearly from surrounding spectral regions before estimating the wavelengths marking the limits of the emission line. The line EW and \tw\ were then directly measured using linear interpolation in the wings of the line. Raw (measurement) uncertainties were estimated by allowing the line boundaries to vary within conservative ranges. Uncertainties on the line EW are on the order of 2 percent, albeit with a ``floor'' uncertainty of 0.1\,\AA\ for the weakest lines. Typical uncertainties on \tw\ range from 5 to 20\,\kms. The EW of several other emission lines, listed in Section\,\ref{subsec:setup}, as well as the EW of the Li\,6707 absorption line, were measured using the same method; the results are reported in Table\,\ref{tab:res}. These are always weak and spectrally unresolved in our spectra.

The \tw\ estimates we obtain are affected by instrumental effects, the most immediate of which is line broadening due to the limited spectral resolution of our data. To correct for this effect, we assume that the convolution of the intrinsic stellar spectrum by the instrumental spectral response readily translates in a quadratic sum of the line widths both for their FWHM and \tw. Implicitly, the underlying hypothesis is that both the intrinsic emission line profile and the instrumental response can be approximated by Gaussian profiles. The spectral profile of isolated lines in the spectra of arc lamps supports this assumption, and we use those to evaluate $W_{\mathrm{10\%}}^{\mathrm{inst}}$. Emission lines in the spectra of TTS, on the other hand, can be complex and depart from simple Gaussian profiles, particularly in the case of \ha. Nonetheless, in the case of relatively narrow lines (\tw\,$\lesssim200$\,\kms), this is a reasonable assumption and thus validates our approach. In cases where the line is intrinsically much broader, the broadening induced by the instrumental resolution is small, and so a departure from a Gaussian line profile has limited consequences on the measured \tw. We then proceed to quadratically subtract $W_{\mathrm{10\%}}^{\mathrm{inst}}$ from all the raw \tw\ estimates to correct them from instrumental line broadening. If the raw estimate of \tw\ is lower than $W_{\mathrm{10\%}}^{\mathrm{inst}}$, precluding this approach, we place an upper limit on the intrinsic \tw\ by generating 100,000 random generalizations of both quantities according to their associated uncertainty and adopt the 99.7 percentile of the resulting distribution of instrument-corrected \tw\ estimates as 3$\sigma$ upper limits. 

% 273 spectra total
% 57 spectra with SNR < 100  (20%)
% 28 spectra with SNR < 75  (10%)
% 11 spectra with SNR < 50  (4%)
%
\begin{landscape}
\begin{table}
\tiny
\caption{Spectral line measurements. The last line for each object provides the average and standard deviation of all \ha\ EW and \tw\ measurements (or the median upper limit on \tw\ in the case of objects where most spectra did not yield a detection), as well as the average and standard deviation of the mean for all detections of the Li\,6707 detections. The SNR listed in the third column is estimated in the 6400--6700\,\AA\ range and is reported on a per resolution element basis. The "accretor" status (14th column) is explained in Sect.\,\ref{subsec:acc_status}.}\label{tab:res}
\begin{tabular}{lccccccccccccc}
\hline
Target & Date & SNR & EW(\ha) & \tw\ & Asym. & \multicolumn{6}{c}{Other Emission Lines EW [\AA]} & Li\,6707 EW & Accretor \\
 &  & & [\AA] & [\kms] & Profile & He\,I\,5876 & [O\,I]\,6300 & He\,I\,6678 & Ca\,8498 & Ca\,8542 & Ca\,8662 & [\AA] & \\
\hline
CoKu\,Tau\,4 & 2009 Oct 27 & 117 & 1.4$^{+\,0.1}_{-\,0.2}$ & $138^{+\,49}_{-\,48}$ & N & \dots & $\geq$\,-0.15 & $\geq$\,-0.15 & \dots & \dots & \dots & 0.59$\pm$0.05 & N \\
 & 2009 Oct 30 & 138 & 0.9$^{+\,0.2}_{-\,0.1}$ & $< 125$ & N & $\geq$\,-0.15 & $\geq$\,-0.15 & $\geq$\,-0.15 & \dots & \dots & \dots & 0.57$\pm$0.05 & N \\
 & 2009 Oct 31 & 213 & 0.9$^{+\,0.1}_{-\,0.2}$ & $< 125$ & N & $\geq$\,-0.15 & $\geq$\,-0.15 & $\geq$\,-0.15 & \dots & \dots & \dots & 0.58$\pm$0.05 & N \\
 & 2009 Nov 07 & 261 & 1.0$\pm$0.2 & \dots & \dots & $\geq$\,-0.60 & $\geq$\,-0.60 & $\geq$\,-0.60 & $\geq$\,-0.60 & $\geq$\,-0.60 & $\geq$\,-0.60 & $\leq$\,0.60 & N \\
 & 2009 Nov 23 & 157 & 0.8$\pm$0.2 & $< 200$ & N & $\geq$\,-0.30 & $\geq$\,-0.30 & $\geq$\,-0.30 & $\geq$\,-0.30 & $\geq$\,-0.30 & $\geq$\,-0.30 & 0.48$\pm$0.10 & N \\
 & 2009 Dec 03 & 76 & 0.8$^{+\,0.1}_{-\,0.2}$ & $< 200$ & N & $\geq$\,-0.45 & $\geq$\,-0.45 & $\geq$\,-0.45 & \dots & \dots & \dots & 0.50$\pm$0.15 & N \\
\hline
\end{tabular}
\end{table}
\end{landscape}

\begin{landscape}
\begin{table}
\tiny
\contcaption{Spectral line measurements.}
\begin{tabular}{lccccccccccccc}
\hline
Target & Date & SNR & EW(\ha) & \tw\ & Asym. & \multicolumn{6}{c}{Other Emission Lines EW [\AA]} & Li\,6707 EW & Accretor \\
 &  & & [\AA] & [\kms] & Profile & He\,I\,5876 & [O\,I]\,6300 & He\,I\,6678 & Ca\,8498 & Ca\,8542 & Ca\,8662 & [\AA] & \\
\hline
IQ\,Tau & 2009 Oct 27 & 86 & 11.6$^{+\,0.4}_{-\,0.6}$ & $373^{+\,14}_{-\,15}$ & Y & \dots & -1.31$\pm$0.05 & -0.22$\pm$0.05 & \dots & \dots & \dots & 0.54$\pm$0.05 & Y \\
 & 2009 Oct 30 & 143 & 8.7$^{+\,0.1}_{-\,0.3}$ & $386^{+\,12}_{-\,14}$ & Y & $\geq$\,-0.15 & -1.21$\pm$0.05 & $\geq$\,-0.15 & \dots & \dots & \dots & 0.55$\pm$0.05 & Y \\
 & 2009 Oct 31 & 210 & 10.5$^{+\,0.5}_{-\,0.6}$ & $368^{+\,14}_{-\,15}$ & Y & -0.23$\pm$0.05 & -1.30$\pm$0.05 & $\geq$\,-0.15 & \dots & \dots & \dots & 0.57$\pm$0.05 & Y \\
 & 2009 Nov 07 & 307 & 12.0$^{+\,0.1}_{-\,0.4}$ & \dots & \dots & $\geq$\,-0.60 & $\geq$\,-0.60 & $\geq$\,-0.60 & $\geq$\,-0.60 & $\geq$\,-0.60 & $\geq$\,-0.60 & $\leq$\,0.60 & Y \\
 & 2009 Nov 23 & 218 & 42.9$\pm$4.0 & $555^{+\,20}_{-\,27}$ & N & -2.34$\pm$0.10 & -1.53$\pm$0.10 & $\geq$\,-0.30 & -3.65$\pm$0.10 & -3.35$\pm$0.10 & -2.88$\pm$0.10 & 0.33$\pm$0.10 & Y \\
 & 2009 Dec 02 & 88 & 24.5$^{+\,0.1}_{-\,0.6}$ & 411$\pm$19 & N & -3.13$\pm$0.15 & -0.64$\pm$0.15 & $\geq$\,-0.45 & \dots & \dots & \dots & 0.48$\pm$0.15 & Y \\
\hline
\end{tabular}
\end{table}
\end{landscape}

\begin{landscape}
\begin{table}
\tiny
\contcaption{Spectral line measurements.}
\begin{tabular}{lccccccccccccc}
\hline
Target & Date & SNR & EW(\ha) & \tw\ & Asym. & \multicolumn{6}{c}{Other Emission Lines EW [\AA]} & Li\,6707 EW & Accretor \\
 &  & & [\AA] & [\kms] & Profile & He\,I\,5876 & [O\,I]\,6300 & He\,I\,6678 & Ca\,8498 & Ca\,8542 & Ca\,8662 & [\AA] & \\
\hline
CX\,Tau & 2009 Oct 27 & 162 & 5.4$\pm$0.1 & $324^{+\,17}_{-\,18}$ & Y & \dots & -0.21$\pm$0.05 & $\geq$\,-0.15 & \dots & \dots & \dots & 0.47$\pm$0.05 & Y \\
 & 2009 Oct 30 & 131 & 4.0$\pm$0.1 & $410^{+\,16}_{-\,23}$ & Y & $\geq$\,-0.15 & $\geq$\,-0.15 & $\geq$\,-0.15 & \dots & \dots & \dots & 0.60$\pm$0.05 & Y \\
 & 2009 Oct 31 & 204 & 5.1$\pm$0.1 & 335$\pm$16 & Y & $\geq$\,-0.15 & -0.21$\pm$0.05 & $\geq$\,-0.15 & \dots & \dots & \dots & 0.62$\pm$0.05 & Y \\
 & 2009 Nov 07 & 150 & 5.7$\pm$0.1 & \dots & \dots & $\geq$\,-0.60 & $\geq$\,-0.60 & $\geq$\,-0.60 & $\geq$\,-0.60 & $\geq$\,-0.60 & $\geq$\,-0.60 & $\leq$\,0.60 & \\
 & 2009 Nov 23 & 228 & 6.2$\pm$0.3 & 341$\pm$26 & Y & $\geq$\,-0.30 & $\geq$\,-0.30 & $\geq$\,-0.30 & $\geq$\,-0.30 & $\geq$\,-0.30 & $\geq$\,-0.30 & 0.36$\pm$0.10 & Y \\
 & 2009 Dec 02 & 119 & 11.1$^{+\,0.2}_{-\,0.3}$ & $419^{+\,30}_{-\,34}$ & N & $\geq$\,-0.45 & $\geq$\,-0.45 & $\geq$\,-0.45 & \dots & \dots & \dots & 0.53$\pm$0.15 & Y? \\
\hline
\end{tabular}
\end{table}
\end{landscape}

\begin{landscape}
\begin{table}
\tiny
\contcaption{Spectral line measurements.}
\begin{tabular}{lccccccccccccc}
\hline
Target & Date & SNR & EW(\ha) & \tw\ & Asym. & \multicolumn{6}{c}{Other Emission Lines EW [\AA]} & Li\,6707 EW & Accretor \\
 &  & & [\AA] & [\kms] & Profile & He\,I\,5876 & [O\,I]\,6300 & He\,I\,6678 & Ca\,8498 & Ca\,8542 & Ca\,8662 & [\AA] & \\
\hline
V807\,Tau & 2009 Oct 27 & 196 & 11.0$\pm$0.4 & $509^{+\,14}_{-\,13}$ & Y & \dots & $\geq$\,-0.15 & $\geq$\,-0.15 & \dots & \dots & \dots & 0.49$\pm$0.05 & Y \\
 & 2009 Oct 30 & 231 & 11.8$^{+\,0.3}_{-\,0.5}$ & $474^{+\,13}_{-\,14}$ & Y & -0.22$\pm$0.05 & $\geq$\,-0.15 & $\geq$\,-0.15 & \dots & \dots & \dots & 0.44$\pm$0.05 & Y \\
 & 2009 Oct 31 & 295 & 15.6$^{+\,0.3}_{-\,0.6}$ & $552^{+\,13}_{-\,16}$ & Y & $\geq$\,-0.15 & $\geq$\,-0.15 & $\geq$\,-0.15 & \dots & \dots & \dots & 0.47$\pm$0.05 & Y \\
 & 2009 Nov 07 & 307 & 19.7$^{+\,0.2}_{-\,0.5}$ & \dots & \dots & $\geq$\,-0.60 & $\geq$\,-0.60 & $\geq$\,-0.60 & $\geq$\,-0.60 & $\geq$\,-0.60 & $\geq$\,-0.60 & $\leq$\,0.60 & Y? \\
 & 2009 Nov 23 & 150 & 8.9$^{+\,0.1}_{-\,0.3}$ & $530^{+\,19}_{-\,20}$ & Y & $\geq$\,-0.30 & $\geq$\,-0.30 & $\geq$\,-0.30 & $\geq$\,-0.30 & $\geq$\,-0.30 & $\geq$\,-0.30 & 0.50$\pm$0.10 & Y \\
 & 2009 Dec 02 & 154 & 6.3$^{+\,0.1}_{-\,0.4}$ & 409$\pm$19 & Y & -0.71$\pm$0.15 & $\geq$\,-0.45 & $\geq$\,-0.45 & \dots & \dots & \dots & $\leq$\,0.45 & Y \\
\hline
\end{tabular}
\end{table}
\end{landscape}

\begin{landscape}
\begin{table}
\tiny
\contcaption{Spectral line measurements.}
\begin{tabular}{lccccccccccccc}
\hline
Target & Date & SNR & EW(\ha) & \tw\ & Asym. & \multicolumn{6}{c}{Other Emission Lines EW [\AA]} & Li\,6707 EW & Accretor \\
 &  & & [\AA] & [\kms] & Profile & He\,I\,5876 & [O\,I]\,6300 & He\,I\,6678 & Ca\,8498 & Ca\,8542 & Ca\,8662 & [\AA] & \\
\hline
FM\,Tau & 2009 Oct 27 & 169 & 58.3$^{+\,0.3}_{-\,1.4}$ & 343$\pm$20 & N & \dots & -0.35$\pm$0.05 & -0.96$\pm$0.05 & \dots & \dots & \dots & $\leq$\,0.15 & Y \\
 & 2009 Oct 30 & 132 & 60.1$^{+\,0.4}_{-\,1.3}$ & 325$\pm$20 & N & -2.00$\pm$0.05 & -0.40$\pm$0.05 & -1.08$\pm$0.05 & \dots & \dots & \dots & 0.16$\pm$0.05 & Y \\
 & 2009 Oct 31 & 155 & 55.9$^{+\,0.2}_{-\,1.6}$ & $309^{+\,20}_{-\,21}$ & N & -2.31$\pm$0.05 & -0.54$\pm$0.05 & -1.03$\pm$0.05 & \dots & \dots & \dots & 0.23$\pm$0.05 & Y \\
 & 2009 Nov 07 & 246 & 59.7$^{+\,0.1}_{-\,1.3}$ & \dots & \dots & -3.64$\pm$0.20 & $\geq$\,-0.60 & -1.22$\pm$0.20 & -1.07$\pm$0.20 & -1.34$\pm$0.20 & -1.04$\pm$0.20 & $\leq$\,0.60 & Y \\
 & 2009 Nov 23 & 193 & 65.0$^{+\,0.4}_{-\,1.5}$ & 366$\pm$31 & N & -4.20$\pm$0.10 & -0.61$\pm$0.10 & -1.45$\pm$0.10 & -1.57$\pm$0.10 & -2.25$\pm$0.10 & -1.59$\pm$0.10 & $\leq$\,0.30 & Y \\
 & 2009 Dec 03 & 152 & 54.9$^{+\,0.6}_{-\,2.4}$ & $376^{+\,30}_{-\,31}$ & N & -4.12$\pm$0.15 & $\geq$\,-0.45 & -1.51$\pm$0.15 & \dots & \dots & \dots & $\leq$\,0.45 & Y \\
\hline
\end{tabular}
\end{table}
\end{landscape}

\begin{landscape}
\begin{table}
\tiny
\contcaption{Spectral line measurements.}
\begin{tabular}{lccccccccccccc}
\hline
Target & Date & SNR & EW(\ha) & \tw\ & Asym. & \multicolumn{6}{c}{Other Emission Lines EW [\AA]} & Li\,6707 EW & Accretor \\
 &  & & [\AA] & [\kms] & Profile & He\,I\,5876 & [O\,I]\,6300 & He\,I\,6678 & Ca\,8498 & Ca\,8542 & Ca\,8662 & [\AA] & \\
\hline
GK\,Tau & 2009 Oct 27 & 134 & 27.9$^{+\,0.7}_{-\,0.8}$ & $524^{+\,13}_{-\,14}$ & Y & \dots & $\geq$\,-0.15 & -0.16$\pm$0.05 & \dots & \dots & \dots & 0.32$\pm$0.05 & Y \\
 & 2009 Oct 30 & 340 & 32.3$^{+\,0.6}_{-\,1.4}$ & $504^{+\,12}_{-\,14}$ & Y & -0.82$\pm$0.05 & $\geq$\,-0.15 & $\geq$\,-0.15 & \dots & \dots & \dots & 0.41$\pm$0.05 & Y \\
 & 2009 Oct 31 & 294 & 24.6$\pm$0.3 & 511$\pm$12 & Y & -0.81$\pm$0.05 & -0.21$\pm$0.05 & $\geq$\,-0.15 & \dots & \dots & \dots & 0.36$\pm$0.05 & Y \\
 & 2009 Nov 07 & 297 & 43.6$^{+\,0.2}_{-\,0.8}$ & \dots & \dots & -1.85$\pm$0.20 & $\geq$\,-0.60 & $\geq$\,-0.60 & -9.78$\pm$0.25 & -11.5$\pm$0.3 & -9.37$\pm$0.25 & $\leq$\,0.60 & Y \\
 & 2009 Nov 23 & 231 & 5.4$\pm$0.1 & $492^{+\,19}_{-\,21}$ & & -0.40$\pm$0.10 & $\geq$\,-0.30 & $\geq$\,-0.30 & $\geq$\,-0.30 & $\geq$\,-0.30 & $\geq$\,-0.30 & 0.33$\pm$0.10 & Y \\
 & 2009 Dec 03 & 110 & 18.5$^{+\,0.1}_{-\,0.4}$ & $474^{+\,19}_{-\,21}$ & N & -1.46$\pm$0.15 & $\geq$\,-0.45 & $\geq$\,-0.45 & \dots & \dots & \dots & $\leq$\,0.45 & Y \\
\hline
\end{tabular}
\end{table}
\end{landscape}

%_ _ _ _ _ _ _ _ _ _ _ _ _ _ _ _ _ _ _ _ _ _ _ _ _ _ _ _ _ _ _ _ _ _ _ _ _ _ _ _ _ 

\subsection{Precision and accuracy of the line measurements}
\label{subsec:simul}

\begin{figure}
\includegraphics[width=\columnwidth]{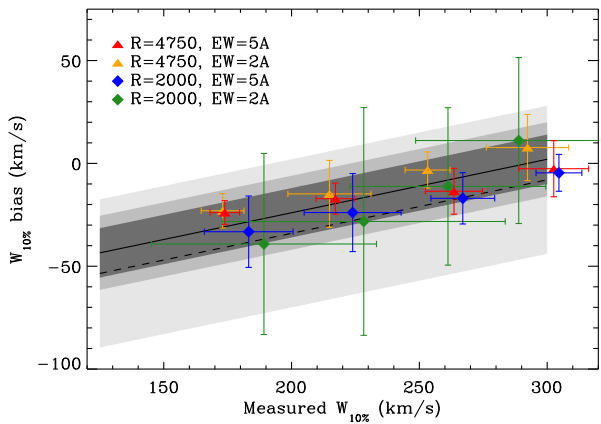}
\caption{Bias on the \ha\ \tw\ (i.e., difference between the intrinsic and derived \tw\ values) derived from our simulations. Symbols with error-bars represent the median and standard deviations of different subsets of our simulations, grouping the two highest and two lowest spectral resolutions for better statistics. The lines represent our best estimate of the bias introduced by the intermediate resolution of our spectra for strong lines in any K- or M-type of star (solid) and weak lines in mid-M targets (dashed). The filled bands represent our 1$\sigma$ confidence interval for three representative cases: strong line in any K- or M-type target observed at the higher resolution (dark gray) and the lower resolution (gray), and weak line in a mid-M target observed at the lower resolution (light gray). Other combinations of line strengths, target spectral type and spectral resolution were also explored but not shown for readability. The dark and light gray bands represent the extreme cases of our entire set of simulations. }\label{fig:simul}
\end{figure}

As a consequence of our limited spectral resolution and spectral sampling, ambiguity in setting the continuum and the presence of photospheric absorption features limit the accuracy of our line measurement. In particular, photospheric features can result in setting the continuum level slightly lower than it ought to. As a result, our methodology leads to slight overestimates of the line EW, albeit at levels that do not exceed our uncertainties. On the other hand, the bias introduced on the \tw\ estimates can be significant. This bias is more severe than the EW one, as it depends on detailed shape of the wings of the line profile, which is particularly delicate to assess. This issue is particularly amplified in cases where some (or all) of the following conditions are met: 1) the emission line is both weak and narrow, leading to a low peak/continuum ratio, 2) the target has a late spectral type, whose spectrum is extremely rich in broad, molecular photospheric features that can affect the continuum evaluation, and 3) the spectral resolution is on the low end of the range used in this study. To help alleviate this last aspect, we have systematically overplotted our highest resolution spectrum of a target over the lower resolution spectra as a template to better evaluate the intrinsic underlying photospheric spectrum. This still leads to larger uncertainties in the case of the lower resolution spectra, but it helped minimizing the systematic bias that could arise from misinterpreting a nearby absorption line for a part of the emission line.

To address the first two issues listed above, we have generated simulated spectra to test the precision and accuracy of our measurement method. The simulations measured the effect of our instruments on spectral templates taken from the MILES stellar library \citep{sanchez06} for 12 stars that cover the mid-K through late-M spectral type range. The simulation proceeded in four steps: constructing the Gaussian response profile of each instrument using the measured instrumental \tw, adding an \ha\ emission line to the template spectrum (which naturally contains a weak absorption line), convolving the two profiles to render the simulated outcome, and applying the adequate instrumental sampling to the result. The emission lines were assumed to have Gaussian profiles centered on \ha\, with our choice of parameters reflecting our interest in lines that are both narrow and weak. Specifically, we produced synthetic spectra with all possible combinations of three EW (0.5, 2 and 5\,\AA) and six \tw\ (ranging from 150\,\kms\ to 400\,\kms\ in 50\,\kms\ increments). With six  instrumental set-ups, we thus built a library of 1080 distinct simulations.
%Each of the resulting 18 profiles per spectral type was then convolved with each instrument's simulated response profile  using the \texttt{FFT\char`_CONVOLUTION} procedure built on IDL’s native FFT functionality, totaling in 1080 simulations. 
The same reduction pipeline described in Section 3 was then applied on each synthetic spectrum to measure the \ha\ \tw, which we then compared with our input parameters to assess the amplitude of the bias introduced by our moderate spectral resolution. 

The overall results from this battery of simulations matches our expectations and can be summarized as follows: 
\begin{itemize}
\item[--] Lower resolution spectra result in larger uncertainties, as well as less constraining upper limits;
\item[--] Emission lines with \tw$\geq$300\,\kms\ are broad enough (and, in practice, strong enough) that there is only a negligible bias, even though it introduces an additional uncertainty of 12--18\,\kms\ depending on the instrumental resolution;
\item[--] Narrow emission lines that have EW$\geq$5\,\AA\ suffer from an essentially linear bias but are strong enough that the details of the underlying continuum (i.e., spectral type) have a negligible impact on this bias, which only depends on \tw\ and instrumental resolution;
\item[--] Narrow and weak emission lines are most susceptible to this bias, although the largest effect introduced by the instrumental correction is to increase the uncertainty associated with the final estimate of \tw. Both the bias and its associated uncertainty are increased at later spectral types due to the richer nature of the photospheric spectrum.
\end{itemize}

Figure\,\ref{fig:simul} illustrates quantitatively the results of our simulations. For an intrinsic \tw\ of 200\,\kms, the amplitude of the bias is on the order of 20\,\kms, with uncertainties ranging from 12 to 30\,\kms, depending on line strength, spectral type and spectral resolution. Therefore, although this bias could have some consequences for the accretion classification of individual spectra, our simulations confirm that the premise of our analysis, namely that we can derive the intrinsic \tw\ of the \ha\ line with our moderate resolution spectra with sufficient accuracy considering the range of expected line widths, is sound. We thus proceed to correct all of our derived \tw\ from this bias using linear fits to the results of our simulations and to increase their associated uncertainties. Typical amplitudes for both effects are illustrated in Figure\,\ref{fig:simul}. As a result of this process, we also had to place upper limits on \tw\ based on the lowest line width that can be confidently retrieved in our simulations. These upper limits range from 125 to 200\,\kms, depending on the instrumental resolution.

We further tested the robustness of the bias correction described above using our own data. We first selected 11 targets spanning the overall ranges of spectral type and \ha\ EW and \tw\ observed in our sample. For each object, we selected a spectrum from our highest resolution setting, and degraded it to the resolution and spectral sampling of our other instrumental set-ups (excluding the low-resolution WHT/ISIS set-up). We then proceeded to measure the raw \tw\ of the \ha\ line, and to correct for both the instrumental broadening and methological bias. Comparing the result \tw\ to those initially measured in the higher resolution spectrum, we find that the results are within 1$\sigma$ of each other in 20 out of 33 cases, and within 2$\sigma$ of each other in all but 2 spectra. These exceptions correspond to the broadest \ha\ lines we tested (\tw$=$550--65-\,\kms) when the spectra are degraded to the lowest resolution (SPM) setting. Overall, these tests confirm that our approach takes properly into account the effects introduced by the medium-resolution used in the survey.

%_ _ _ _ _ _ _ _ _ _ _ _ _ _ _ _ _ _ _ _ _ _ _ _ _ _ _ _ _ _ _ _ _ _ _ _ _ _ _ _ _ 

\subsection{Comparison to previous measurements}

\begin{figure*}
\includegraphics[width=0.49\textwidth]{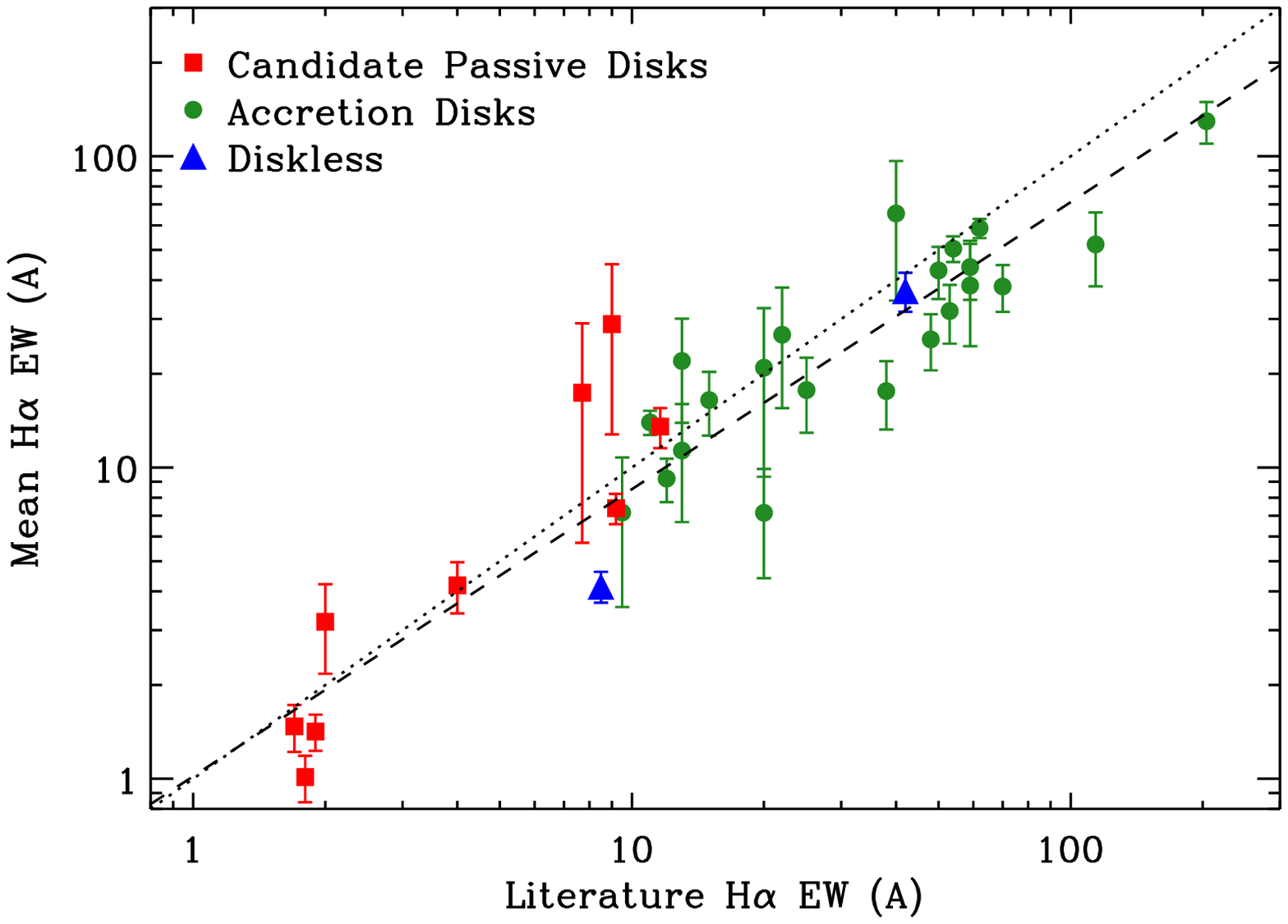}
\includegraphics[width=0.49\textwidth]{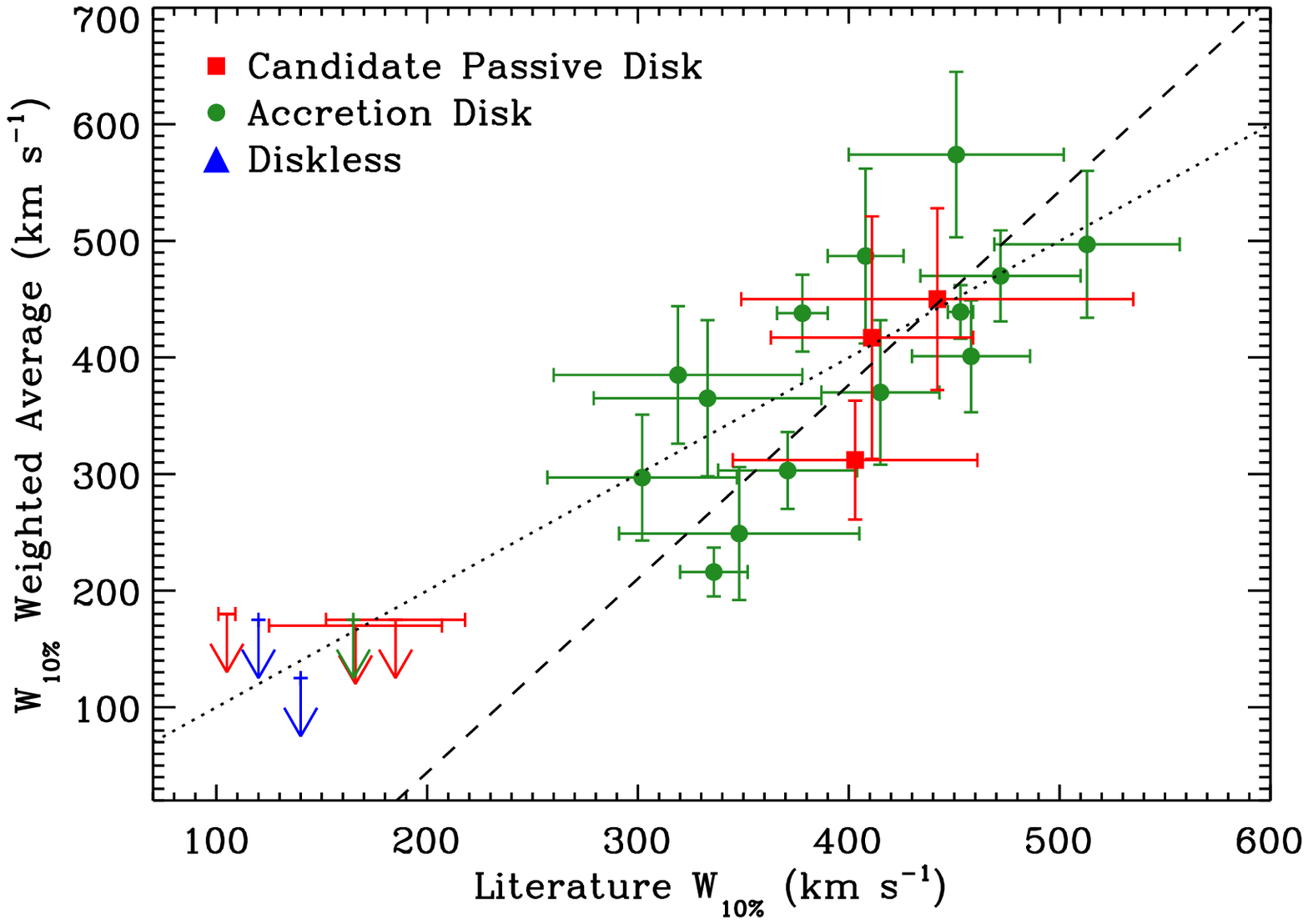}
\caption{{\it Left:} Comparison of the mean \ha\ EW from our survey to previously published estimates (see Table\,\ref{tab:sample}) for all objects in our sample. Vertical uncertainties represent the standard deviation from all measurements of a given object. The dotted and dashed lines mark the identity relation and an unconstrained least squares linear fit to all objects, respectively. {\it Right: } Comparison of the \tw\ \ha\ measurements from our survey to previously published estimates \citep{white03, nguyen12}. From our monitoring campaign, we plot the weighted mean and the largest of 1) the standard deviation of all measurements for a given target, or 2) the mean associated to the weighted mean. Symbols are as in Figure\,\ref{fig:sample}. For objects with too few estimates of \tw, we show the median of all measurements and upper limit from our survey. The dotted line marks the identity relation to guide the eye whereas the dashed line is the result of a linear fit to objects with \tw\,$\geq 250$\,\kms. }\label{fig:compar_w10}
\end{figure*}

Before combing our datasets for individual objects with an unexpected accretion state, we first present an overview of the results by comparing our measurements of both EW and \tw\ of the \ha\ line to those listed in the literature \citep{white03, nguyen09}. For each star in our sample, we compute the uncertainty-weighted average of both quantities. To these mean quantities, we associate an ``uncertainty'' which is the largest of 1) the dispersion of all values, and 2) the uncertainty on the weighted average. In practice, for all EW and most \tw, the former is larger than the latter. For some objects, most (or all) of our spectra led to an upper limit on \tw. In these cases, we assigned an upper limit to the object which is the median of all measurements and upper limits. Figure\,\ref{fig:compar_w10} shows the comparisons of our average measurement with literature data.

Some objects show significant departures from the unity relationship, which are indicative of significant long-term variability. This effect could at least partially explains some of the apparently passive disk systems as the \ha\ EW has been underestimated in past studies of some of our targets. Furthermore, our EW values are lower by $\approx$15 per cent on average relative to literature measurements, which we suspect is due to the more modest spectral resolution of those earlier EW-driven studies, which leads to biased estimates of the continuum level. Nonetheless, the strong correlations between our measurements and previous ones are statistically very significant (at the $\geq 3\sigma$ and $\geq 7\sigma$ levels for \tw\ and EW, respectively, using either the Spearman or Kendall tests). The slopes of the linear fits are 0.92$\pm$0.05 and 1.66$\pm$0.31 for the EW and \tw\ measurements respectively (including only objects with \tw$\geq$250\,\kms\ for the latter fit), i.e., both fits are consistent with the 1:1 relationship within $\approx2\sigma$. Since most of the literature data used to make these plots have been obtained several years ago, we thus conclude that the average \ha\ EW and \tw\ of TTS do not dramatically change over such timescales. For instance, the median difference between our average \tw\ value and that from the literature is -5\,\kms, with a dispersion of 67\,\kms. 

%_ _ _ _ _ _ _ _ _ _ _ _ _ _ _ _ _ _ _ _ _ _ _ _ _ _ _ _ _ _ _ _ _ _ _ _ _ _ _ _ _ 

\subsection{Timescales of line variability}

\begin{figure*}
\includegraphics[width=0.49\textwidth]{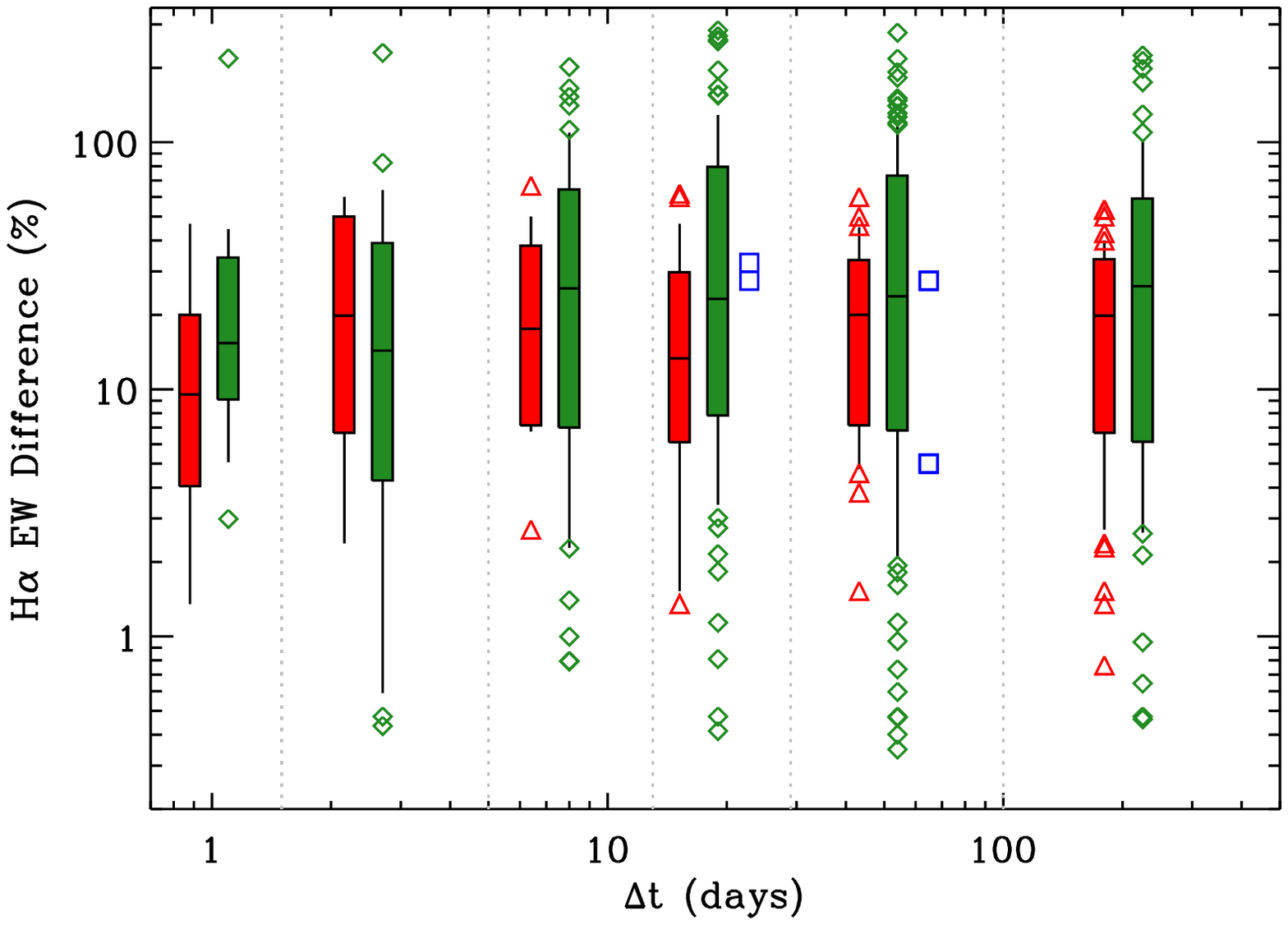}
\includegraphics[width=0.49\textwidth]{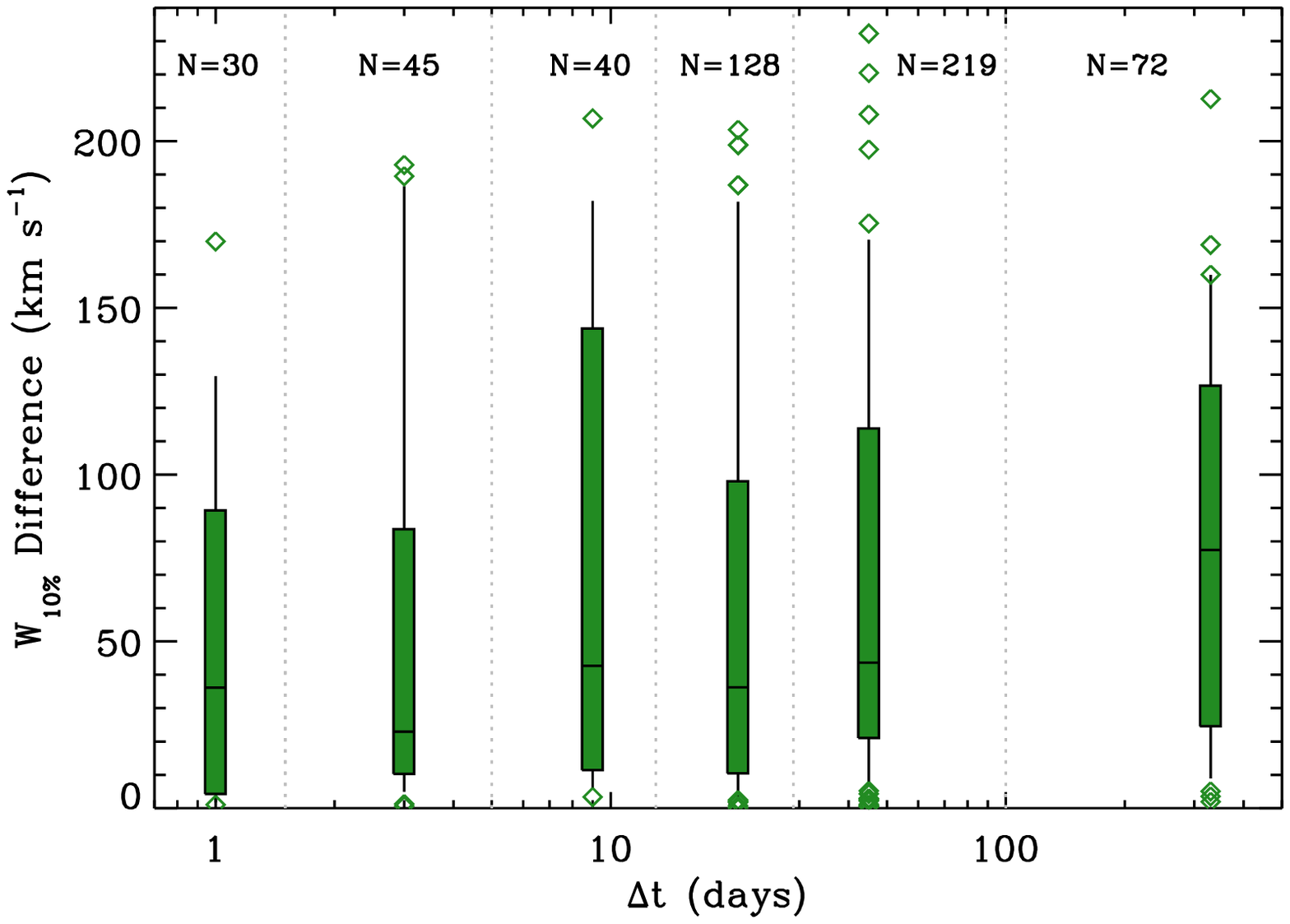}
\caption{{\it Left:} Difference in EW (relative to the median EW over all epochs) as a function of time interval for all objects in our sample. Green diamonds, red triangles and blue squares represent accreting, non-accreting and disk-less objects (only UX\,Tau\,C belongs to the latter category), respectively; they are separated horizontally for display purposes only. To minimize confusion, we binned the data using the vertical gray lines to mark the bin separations, which correspond to natural breaks in our temporal coverage. In each bin, we show the median, 68 and 90 percentiles as the horizontal segment, rectangular box and whiskers, respectively. The most extreme outliers are shown as individual datapoints. In cases where less than 10 objects were available in a bin, we plot all individual measurements as separate symbols instead. {\it Right:} Difference in \tw\ as a function of time interval for all accreting objects in our sample. We have too few \tw\ detections for non-accreting and disk-less objects (4 objects with 2 measurements each). Symbols are the same as in the left panel. The number of pairs of measurements included in each bin are indicated at the top. }\label{fig:ha_time}
\end{figure*}

As mentioned above, the dispersion of values we find for both the \ha\ EW and \tw\ is much larger than the uncertainty associated with each individual measurement. Focusing on \tw, the largest epoch-to-epoch fluctuation is significant at the $\geq 2 (3) \sigma$ level for 22 (17) out of the 27 objects with 2 or more measurements (i.e., excluding upper limits). The most significant differences exceed the 9$\sigma$ confidence level. While the 5 remaining systems have non-significant variations ($\lesssim 1 \sigma$), we only obtained 2 or 3 \tw\ estimates for each of them. Furthermore, 4 of these 5 objects are not accreting (see Sect.\,\ref{subsec:acc_status}), for which we expect the line profile to be more stable. Overall, the large dispersion of \tw\ estimates for most targets illustrates the long-established strong variability of the line profile on timescales as short as 1\,d. Nonetheless, the dispersion in \tw\ that we measure is comparable to previous dispersion estimates for the same targets \citep[e.g.,][]{nguyen09}, with a median ratio of 1.4 and with most systems below a ratio of $\approx$2. This suggests that the amplitude of fluctuations of \tw\ does not increase dramatically beyond the $\approx1$\,yr timescale of individual studies.

The sampling of our monitoring campaigns allows us to probe a wide range of timescales, from 1\,d to 1\,yr, thus complementing the above analysis which addresses multi-year timescales. Specifically, we evaluate the epoch-to-epoch variation of both quantities for each pair of observations of a given target, leading to up to 55 independent EW or \tw\ differences per source. We then group all the resulting differences by bins of time delay: $\approx$1\,d, 2--4\,d, 6--12\,d, 14--28\,d, 30--60\,d and 300--360\,d. These bins correspond to natural breaks in our campaign sampling. They also allow an inspection of the degree of line variability on timescales of days, weeks, months, and 1\,year. Because the EW values span such a large range over our entire sample (from $\lesssim$1\,\AA\ to $\geq$100\,\AA), and because accreting TTS have systematically much larger EWs than non-accreting TTS, we focus on the relative differences in EW.

Figure\,\ref{fig:ha_time} summarizes the variability of both the \ha\ EW and \tw\ over all timescales covered in this study. Here we group objects based on their accretion status, which is derived in Section\,\ref{subsec:acc_status}. The first immediate observation is that accreting TTS display much wider amplitude of variability in EW, with epoch-to-epoch variations sometimes exceeding a factor of 3 change in line intensity whereas non-accreting TTS do not exceed 70 per cent epoch-to-epoch fluctuations. Combining all timescales, a Wilcoxon rank-sum test confirms at the 5.4$\sigma$ confidence level that accreting TTS experience a higher median degree of variability than non-accreting objects. This difference in amplitude of variability between accretors and non-accretors is statistically significant on timescales longer than 1 month, and there is marginal evidence that it may extend down to timescales of 1 week. Our observing cadence results in a poorer sampling of shorter timescales. On the other hand, non-accreting and disk-less objects share similar variability properties over all timescales within uncertainties. %1.5sigma difference 

Only accreting TTS yield enough \tw\ measurement to probe the variability of this quantity: we could only measure \tw\ (rather than obtaining an upper limit) for two epochs for only 4 of our 11 non-accreting or disk-less targets. We thus focus on accreting TTS when studying \tw\ variability. First of all, we note that the amplitude of variations in \tw\ can be large: the 95 percentile exceeds 150\,\kms\ over all timescales longer than 1\,d, compared to a median of 410\,\kms\ over all accreting targets. However, both the upper envelope to the distribution and its 95 percentile are independent on timescale. A uniform degree of variability over all timescales longer than 1\,d is consistent with the results of \citet{nguyen09} and \citet{costigan14}. Our sampling does not provide sub-day coverage, for which these authors found that the variability amplitudes are reduced. 

On the other hand, we notice a gradual increase in the median \tw\ difference with longer timescale: it rises from 34, to 43 and to 77\,\kms\ for timescales of less than 1\,month, 1--2\,months, and 1\,yr. Computing similar median differences in \tw\ from the data shown in \cite[][their Figure\,3]{nguyen09}, we find median differences of 34, 46 and 27\,\kms\ over the same timescales, respectively. On timescales up to 2\,months, both studies yield consistent results, confirming the robustness of the gradual increase. On the other hand, the much reduced difference in \tw\ on a timescale of 1\,yr in the \cite{nguyen09} is opposite the results from our survey. However, we note that the 2010 part of our campaign was mostly focused on non-accreting TTS, so that very few measurements for accreting TTS contribute to the long-timescale bin (9 individual \tw\ measurement covering 5 targets). As a result, the influence of a handful of lower-than-average \tw\ is amplified by our computing differences from all possible pairs. Of the 5 objects contributing to this bin, only UX\,Tau\,A's lone 2010 observations represent a significant outlier compared to its 2009 observations, but that particular spectrum is characterized by a strong inverse P\,Cygni profile that accounts for the low apparent \tw\ (see Figure\,\ref{fig:spec_ex}). We thus do not consider that the apparent contradiction on the 1-yr timescale between our results and those of \cite{nguyen09} represents a meaningful difference.

%_ _ _ _ _ _ _ _ _ _ _ _ _ _ _ _ _ _ _ _ _ _ _ _ _ _ _ _ _ _ _ _ _ _ _ _ _ _ _ _ _ 

\subsection{Accretion status}
\label{subsec:acc_status}

\begin{figure*}
\includegraphics[width=0.49\textwidth]{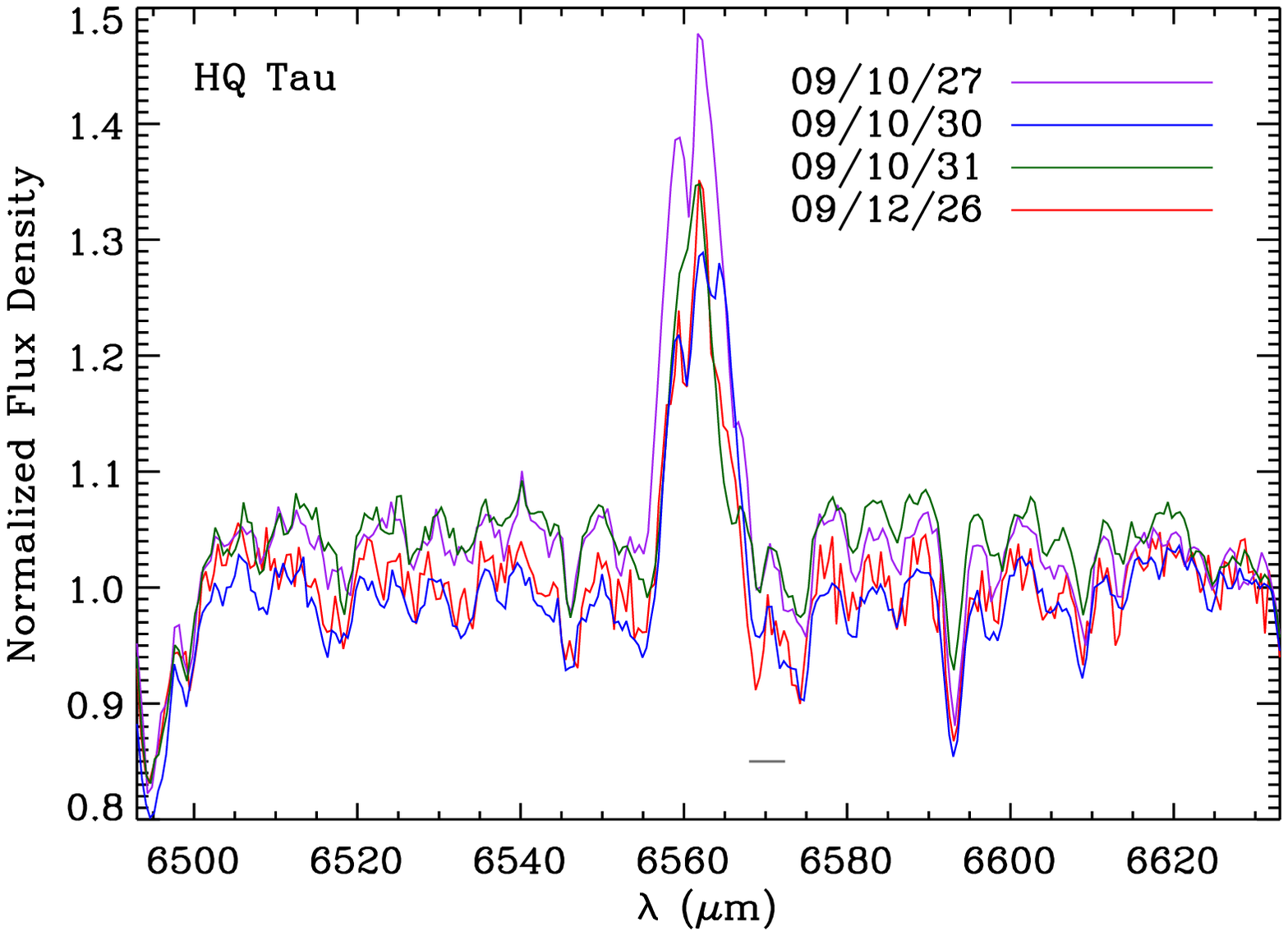}
\includegraphics[width=0.49\textwidth]{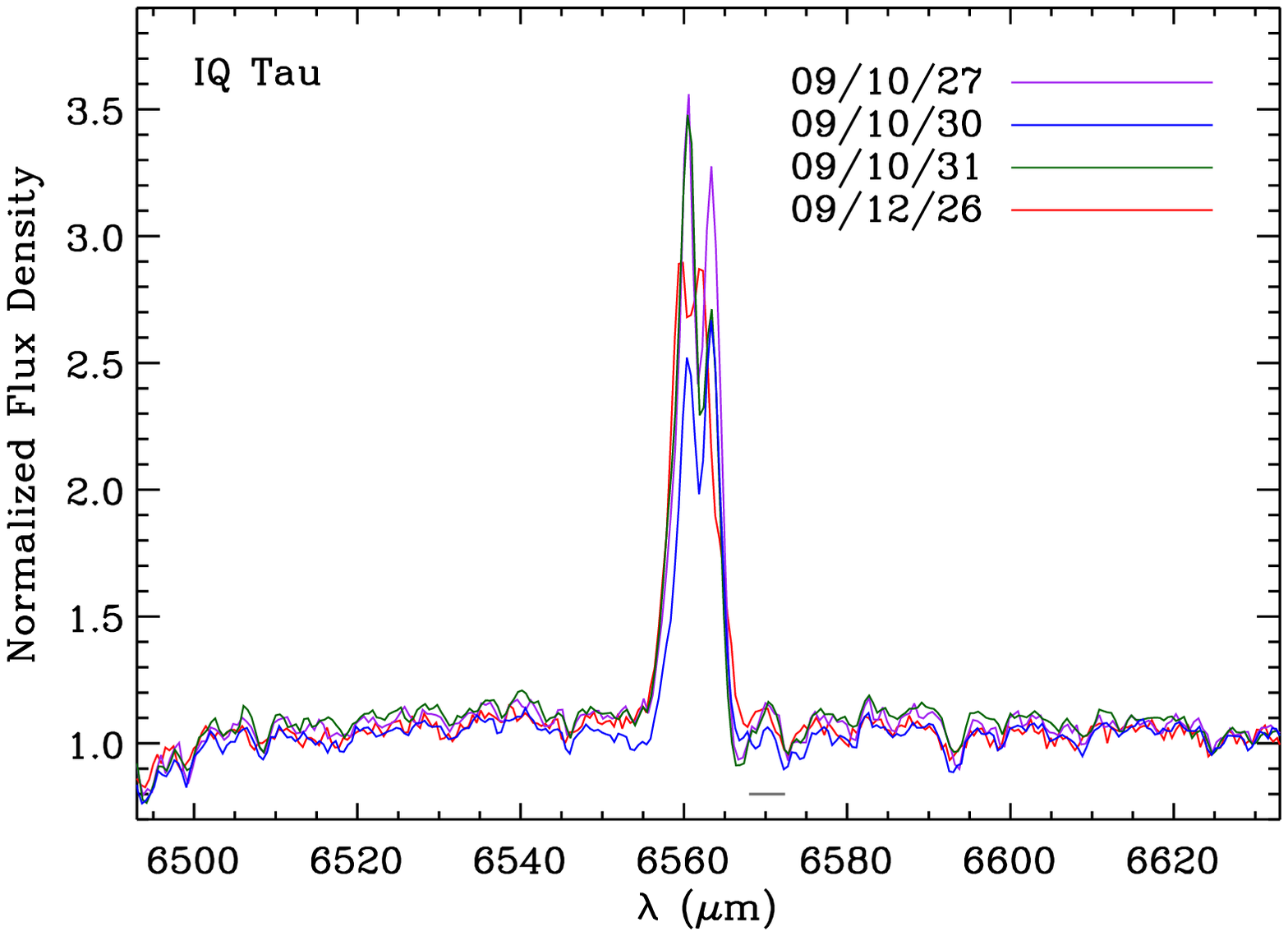}\\
\includegraphics[width=0.49\textwidth]{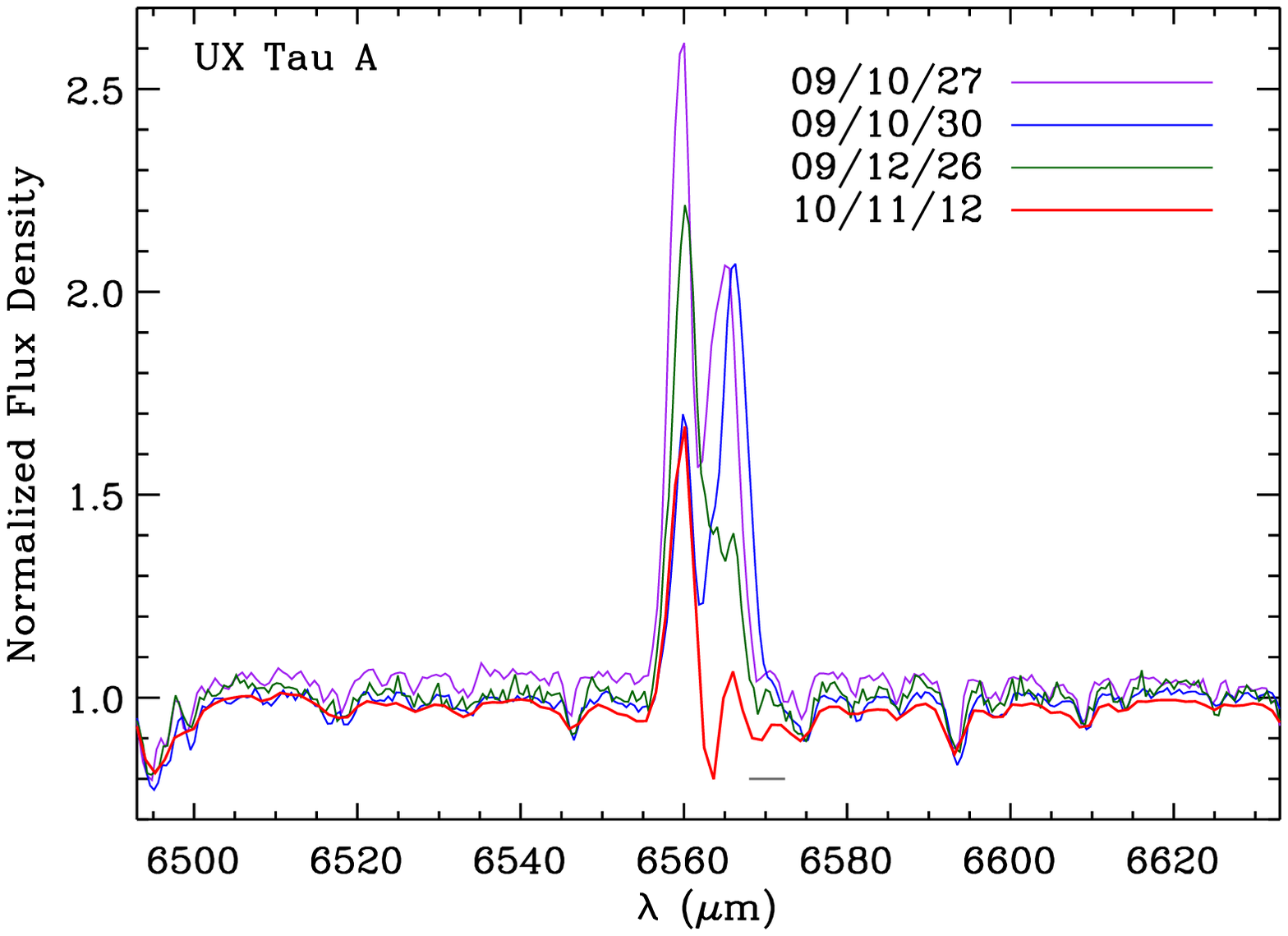}
\includegraphics[width=0.49\textwidth]{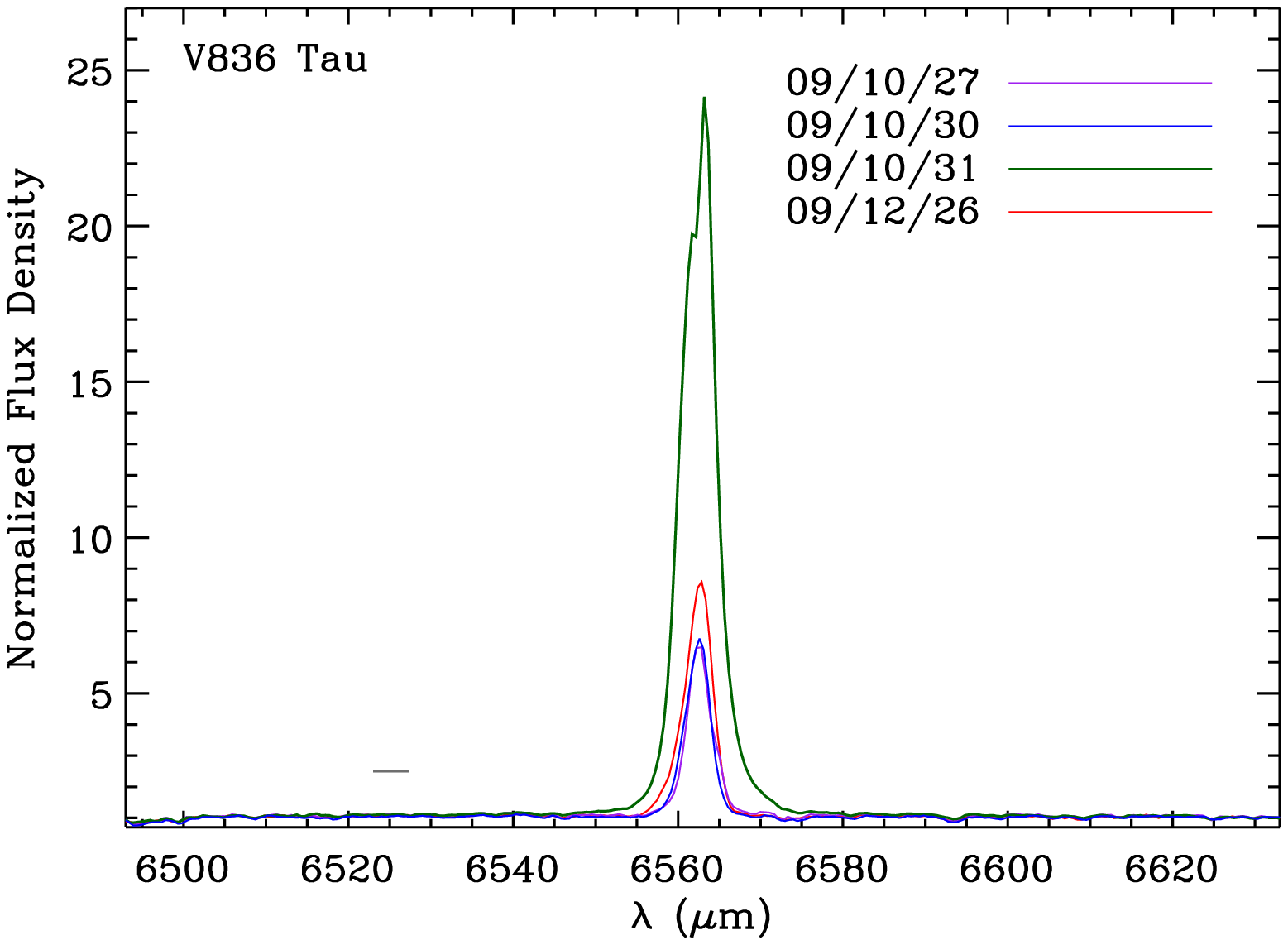}
\caption{Example of continuum-normalized spectra for remarkable sources in our sample, illustrating the large variability in line profile and/or strength. For reference, the horizontal gray bar represents 200\,\kms\ at the wavelength of \ha. }\label{fig:spec_ex}
\end{figure*}

The next step in our analysis consists in determining which objects were actively accreting during our observations and when. Since we want to test the robustness of \ha-based criterion, we must rely as much as possible on other tracers in our spectra. Specifically, for each spectrum of each object, we classified the object as accreting if at least one the following conditions was fulfilled: 1) the \ha\ line profile show unambigous asymmetry (double peak, broad red- or blue-shifted "shoulder, P\,Cygni profile), 2) either of the two He\,I lines is detected in emission, 3) at least one line of the Ca triplet is detected in emission, and/or 4) the EW of Li\,6707 absorption line is significantly different from the weighted average of all other detections for that source. Because of the lack of spectro-photometric quality and the limited set of template spectra from non-accreting objects, our data do not provide a robust estimates of the veiling, so we use the latter condition as a proxy for measuring a change in continuum veiling. 

If neither of these four conditions is fulfilled, it is tempting to consider as non-accreting at that epoch. However, there are reasons to adopt a more cautious approach. First of all, some of our spectra do not include the He\,I\,5876 nor the Ca triplet, which are among the strongest lines in accreting TTS. Furthermore, even these lines remain undetected in objects whose accretion status is established by UV/blue excess emission \citep[e.g.,][]{alcala16}. Similarly, when the \ha\ line profile is relatively narrow, modest-to-weak profile asymmetries may remain undetected at our spectral resolution. Finally, our precision in measuring the Li\,6707 EW (typically ranging from 0.05 to 0.15\,AA) is only sufficient to detect strong variability in accretion-induced veiling. Indeed, veiling in the red optical is undetected even in confirmed accretors \citep[e.g.,][]{frasca17}. As a result, it is possible that a spectrum in which neither of the four conditions listed above is fulfilled is nonetheless associated with an accreting TTS. To solve for this ambiguity, we adopted the following method: if neither condition is fulfilled in any spectrum of a given target, the object is considered as non-accreting throughout our survey. We caution that it remains possible that these objects are accreting at low enough accretion rates that no tracer is detectable in our spectra, but we consider unlikely that this is true for all epochs on given target. If, on the other hand, the object was distinctively accreting at some epochs, but remains ambiguous at others, the accretion status for the latter is based on the \ha\ EW and \tw: the objects is considered "probably accreting" (marked as "Y?" in the 9th column of Table.\,\ref{tab:sample}) at epochs when the line was at least as strong and/or wide as it was at clearly accreting epochs. 

In our sample, there are 11 objects which are unambigously accreting at all epochs. Unsurprisingly, all but one of them are part of our control sample of accretors. The lone exception is IQ\,Tau, which is part of candidate passive disk subsample. As illustrated in Fig\,\ref{fig:spec_ex}, the \ha\ line profile for this object is markedly double peaked at multiple epochs, leaving no doubt about its accretion status. Conversely, 5 objects appear to be never accreting in our survey. Four of them are part of our candidate passive disk subsample (CoKu\,Tau\,4, FW\,Tau, RX\,J0432+1735 and V819\,Tau) and thus our data confirm that those objects show no sign of accretion over timescales ranging from 1\,d to at least 1\,yr.

A group of 14 targets show a common behavior of unambiguously accreting at multiple epochs and "probably accreting" at all other epochs. For lack of contradictory evidence, we surmise that these objects were continuously accreting throughout our survey. This group includes two candidate passive disks. HQ\,Tau has a consistently weak but broad, and often double-peaked, \ha\ line. It is therefore a textbook example of a TTS whose line profile is strongly affected by strong self-absorption, resulting in an uncharacteristically low EW (Figure\,\ref{fig:spec_ex}). UX\,Tau\,A displays a similar behavior, with a marked inverse P\,Cygni profile. The other candidate passive disk which we find to be actually accreting is V836\,Tau. This is an example of a system for which at least one historical EW measurement is much weaker than those measured in our spectra (18--75\,\AA). The original measurement is from \cite{mundt83}, who noted significant line variability. Intense line variability is evident in our data for this source, with a strong accretion outburst observed on 2009 Oct 31 (Figure\,\ref{fig:spec_ex}). In the absence of the original spectra from \cite{mundt83}, it is unclear whether the low EW epochs are the consequence of line self absorption or an actual pause in the accretion flow on the central star. In any event, the detection of He\,I emission lines unambigously establishes this object as accreting in our survey.

One unexpected member of this group is MHO\,4, a disk-less object which is not expected to undergo accretion at all. Yet, either of the He\,I lines is detected at 4 epochs, a fact that was also noted in the discovery study of \cite{briceno98}. This leads to an ambiguous interpretation for this source. Either it possesses a circumstellar disk that has so far escaped detection even at mid-infrared wavelengths \citep{luhman10}, or the emission lines instead originate in the surrounding cloud and this extended emission is not perfectly subtracted in usual sky subtraction in long-slit spectra. In the absence of conclusive evidence pointing either way, we consider this object has a likely non-accretor but do not include it in any statistical test to avoid any bias.

FP\,Tau is the only object in our survey that has one epoch with an apparently weak \ha\ line compared to the other epochs and no other accretion tracer. While it possible that this object experienced a short pause in accretion, we note that its \ha\ line is only marginally weaker than at other epochs, and thus we consider it more likely that it is continuously accreting. We list its accretion status as "Y?" in Table\,\ref{tab:sample}.

Finally, two objects (CZ\,Tau and V410\,Tau\,X-ray\,6, both candidate passive disks) would be unambiguously characterized as non-accretors in our survey except for the fact that we detect significant [OI] emission at multiple epochs. This is puzzling, as the line is usually a tracer of outflow, which itself requires active accretion onto the central star. Therefore the origin of this emission line in these objects is uncertain. Besides the possibility of contamination by improperly corrected atmospheric airglow and/or background emission from the surrounding molecular cloud, we note that CZ\,Tau is an unresolved and unusual binary system, possibly accounting for the line emission pattern (see Sect.\,\ref{subsec:discuss_passive}). Based on all other emission lines, we classify both objects as "probably not accreting."

%________________________________________________________________

\section{Discussion}
\label{sec:discuss}

%_ _ _ _ _ _ _ _ _ _ _ _ _ _ _ _ _ _ _ _ _ _ _ _ _ _ _ _ _ _ _ _ _ _ _ _ _ _ _ _ _ 

\subsection{Passive disks and flickering accretion in Taurus}
\label{subsec:discuss_passive}

\begin{figure*}
\includegraphics[width=0.49\textwidth]{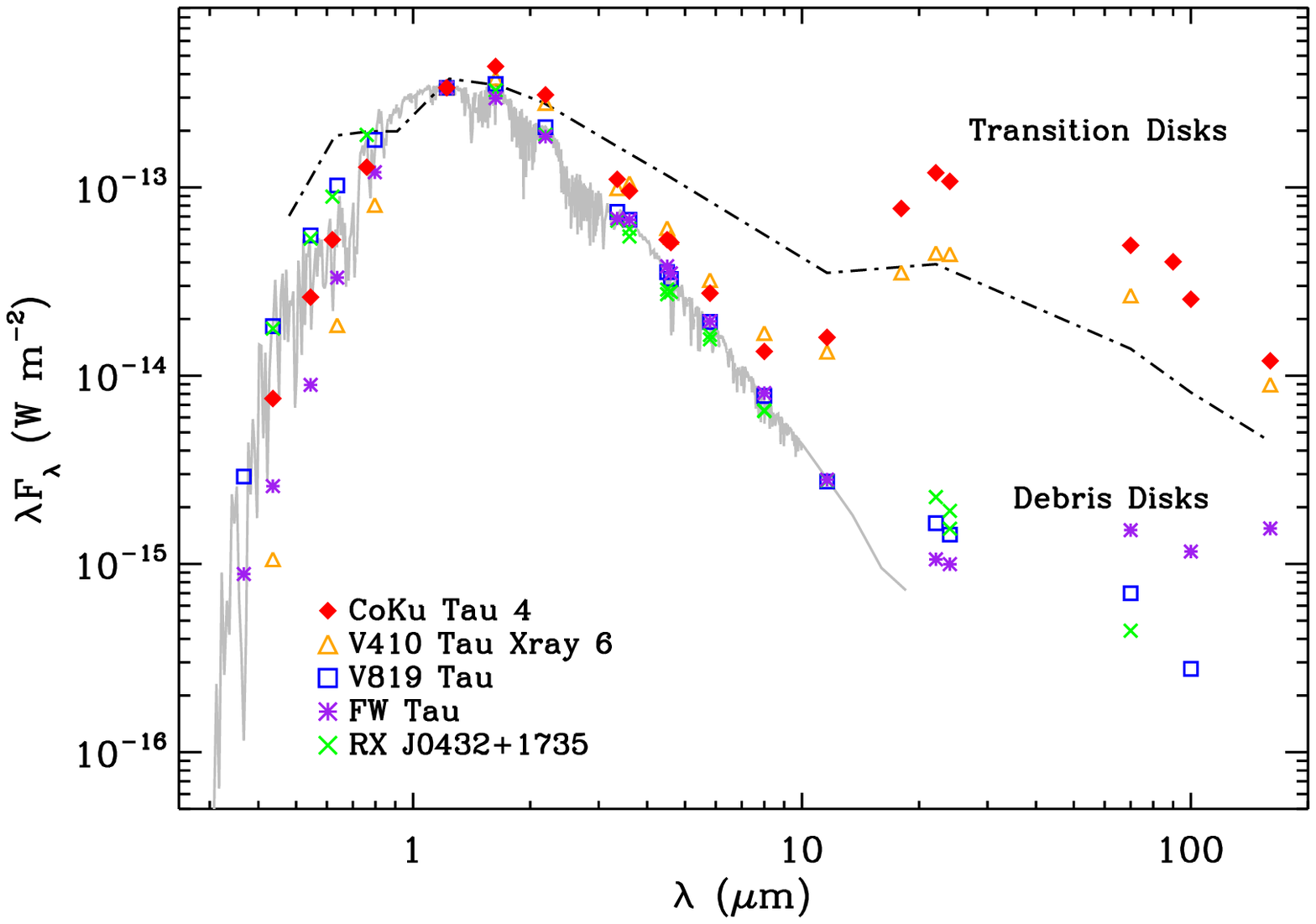}
\includegraphics[width=0.49\textwidth]{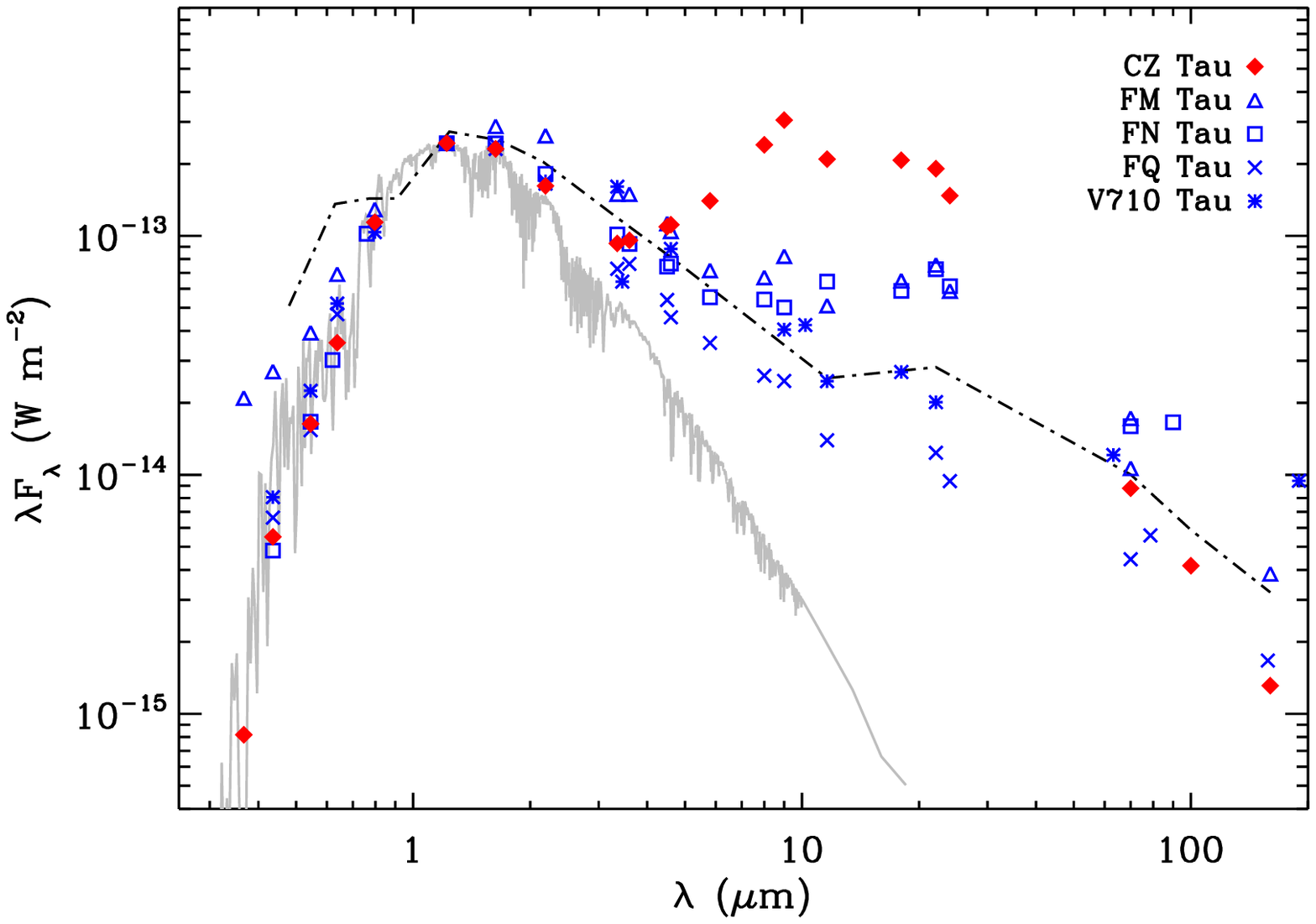}
\caption{{\it Left:} SEDs for all objects possessing a circumstellar disk but identified as non-accretors in our survey, with the exception of CZ\,Tau. The gray spectrum is 3200\,K NextGen stellar model \citep{allard97}, while the dot-dashed curve is the SED for the "median" Taurus Class\,II source \citep{mathews13}. All SEDs are normalized to their $J$ (1.25\,$\mu$m) flux for easier comparison. {\it Right:} SED for CZ\,Tau compared to the 4 targets in our sample that possess a disk and have a spectral type within a half subclass of that CZ\,Tau. Curves and normalization are as in the left panel. As expected, the accreting Class\,II sources scatter around the median SED, whereas CZ\,Tau lies an order of magnitude higher in the mid-infrared.}\label{fig:passive_seds}
\end{figure*}

Overall, we have identified 7 non-accreting objects in our sample. In particular, 6 out of the 9 candidate passive disks are confirmed to be non-accretors. The remaining 3 sources in that sample either masqueraded as weak \ha\ emitters whose broad line revealed their true nature (HQ\,Tau), or display a strongly variable line strength, with peak-to-trough ratios exceeding a factor of 4 (IQ\,Tau and V836\,Tau). As it turns out, the historical EW measurements of these latter two sources were obtained at epochs of weak line emission, accounting for their initial classification. Unfortunately, these historical measurements do not have corresponding \tw\ estimates, so it remains impossible to decide whether they were accreting or not at that time. The only robust statement we can make is that those objects were consistently accreting throughout our monitoring campaign. Weeding out such objects from our initial sample of candidate passive disks was one of the goals of our project.

Taken at face value, our campaign confirms the passive disk nature of the 6 objects with a confirmed disk but no sign of accretion. However, all but two of these objects are instead transition disks \citep[CoKu\,Tau\,4 and V410\,Tau\,X-ray\,6;][]{luhman10} or have extremely weak IR excesses and, thus, best characterized as young debris disks \citep[V819\,Tau and RX\,J04328+1735;][]{furlan09, wahhaj10, hardy15}. None of these objects has an infrared excess out to $\approx$10\,$\mu$m (Figure\,\ref{fig:passive_seds}), indicating an absence of dust in the inner $\gtrsim$1\,AU. The apparent absence of accretion on the central star is therefore not surprising. 

The remaining two objects are cases in which interpretation is muddled by multiplicity. FW\,Tau is a triple system in which the lone circumstellar disk is associated with a substellar component \cite{kraus15} whereas the optical spectra are dominated by the close pair consisting of two disk-less WTTS. CZ\,Tau is a close (0\farcs3) binary system whose SED is not typical of Class\,II objects. Figure\,\ref{fig:passive_seds} shows that, compared to other Class\,II sources with similar spectral types in our sample, CZ\,Tau is characterized by a remarkably strong mid-emission. Indeed, the bolometric luminosity of the system is actually dominated by the mid-infrared emission from the system rather than the near-infrared. While such a double-hump shape is often associated with edge-on disk systems, CZ\,Tau is not under-luminous as is typical for these special systems. Instead, CZ\,Tau is a member of the small category of "infrared companions" \cite{chelli88, koresko97}, as further confirmed by the strongly wavelength-dependent flux ratio observed by \cite{mccabe06}. Therefore, this is another case where the optical light, which indicates no accretion, is associated with a different component than the infrared emission, which reveals the presence of a disk. 

In summary, our search for permanently non-accreting full-fledged disks has led to a null result. While the existence of passive disks has long been suggested, it appears that sufficient scrutiny of each individual objects rules out this particular configuration, at least in the Taurus star-forming region. Instead, the 5--10 per cent occurrence rate suggested by large surveys \citep[e.g.,][]{mccabe06, hernandez14, petterson14} probably consists of a combination of transition disks,  debris ("anemic") disks, and weak-but-broad-line impostors in these regions. 

On the other hand, the (presumably less evolved) pre-transitional disks in our sample, LkCa\,15 and UX\,Tau\,A, which are characterized by a mid-infrared deficit but significant near-infrared excess, were accreting at all epochs. In other words, disks which extend inwards to the immediate vicinity of the star always support ongoing accretion on the central objects. In turn, this implies that the disk clearing process is essentially simultaneous, in the sense that dust is cleared out as soon as the accretion flow is halted (or vice versa). From a physical standpoint, the implication is that it is virtually impossible to hold a significant amount of material from accreting on the central star, suggesting that viscous evolution of the inner disk always dominates the dynamical state of the inner disk, irrespective of the how the accretion flow is initially sustained.

A second goal of the survey was to identify objects whose accretion was flickering, i.e., TTS that appear accreting at times but not always. Focusing first on objects whose \ha\ EW temporarily crosses the spectral type-dependent threshold for accretion, a method that has been used in the past to suggest flickering accretion \citep[e.g.,][]{murphy11, riviere15}, we found 8 such objects in our survey. Of these, six have a broad line as measured by their \tw\ and thus the apparently weak line is induced by self-absorption rather than by a pause in accretion. The remaining object is DN\,Tau whose \ha\ EW hovers around the threshold, but has multiple detections of the He\,I\,5876 emission line. The accretion status of this object was independently confirmed by spectropolarimetry \citep{donati13}. We interpret this objects as an example of the ambiguous nature of any \ha-only criterion. 

In conclusion, we did not identify any object that unambiguously experienced a temporary pause in accretion during the course of our monitoring survey, suggesting that such events are rare. Limiting the analysis to the 24 accreting objects with multiple observations in our survey, we obtained a total of 158 spectra. This implies an upper limit of 2.2 per cent (95 per cent confidence level) for the duty cycle of accretion "gaps" in TTS, assuming all stars undergo such events. No definitive example of a complete halt to the accretion flow on a young star has been identified yet, although large variations in accretion rates, traced both by line strength and by indirect evidence of the accretion column passing in front of the star, have been recorded in the case of AA\,Tau \citep{bouvier03, bouvier08}. The inherent stochastic variability of this and similar system precludes determining precisely the duty cycle of such phenomenon. However, the stringent upper limit on the duty cycle derived here may be so low that it would make it unlikely to detect such an event for any single object. In turn, this would imply that only objects in a particular state of evolution, possibly those close to disk dissipation, can experience accretion "gaps".

%_ _ _ _ _ _ _ _ _ _ _ _ _ _ _ _ _ _ _ _ _ _ _ _ _ _ _ _ _ _ _ _ _ _ _ _ _ _ _ _ _ 

\subsection{On the use of \ha\ as an accretion discriminator}
\label{subsec:discuss_sptdep}

\begin{figure*}
\includegraphics[width=0.49\textwidth]{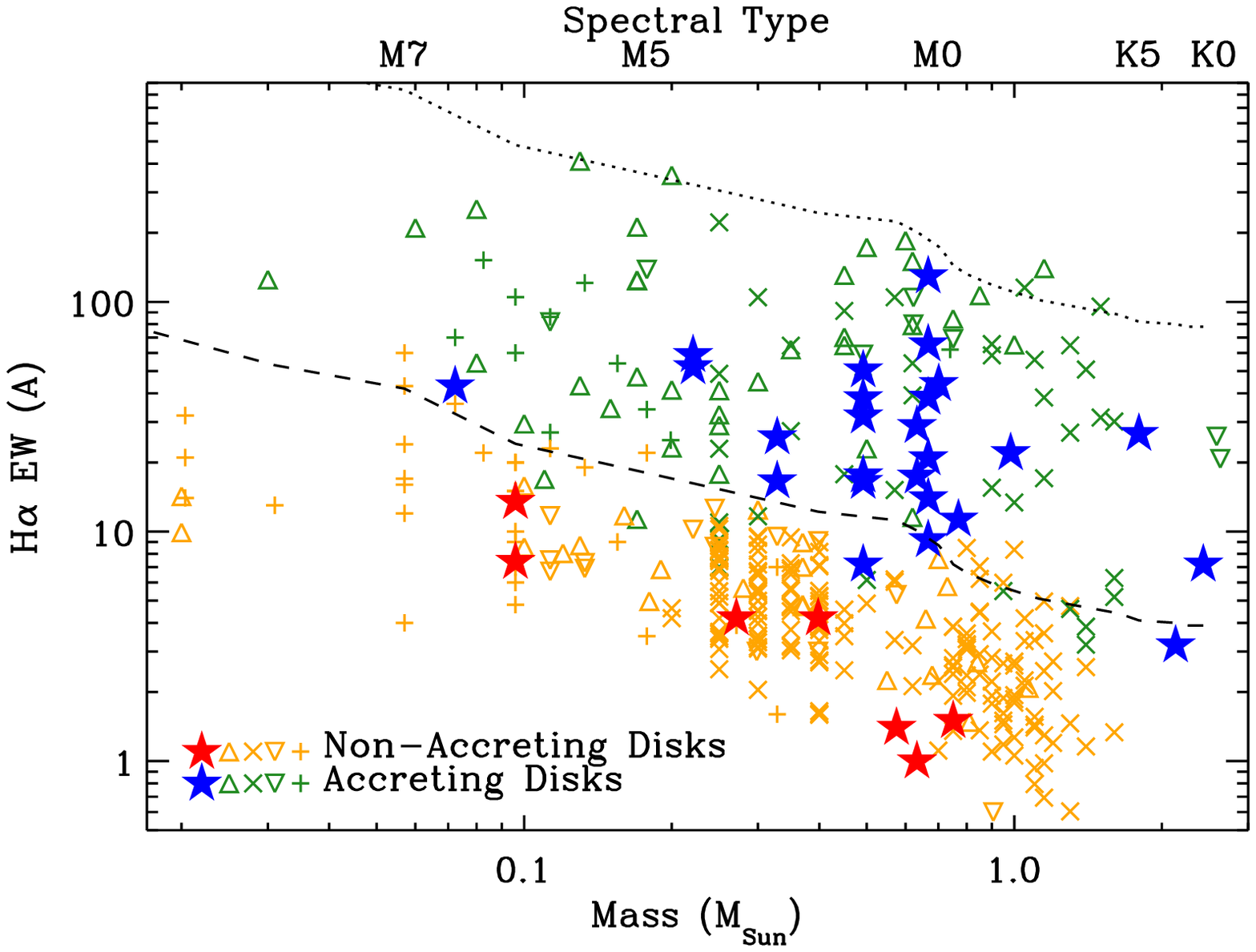}
\includegraphics[width=0.49\textwidth]{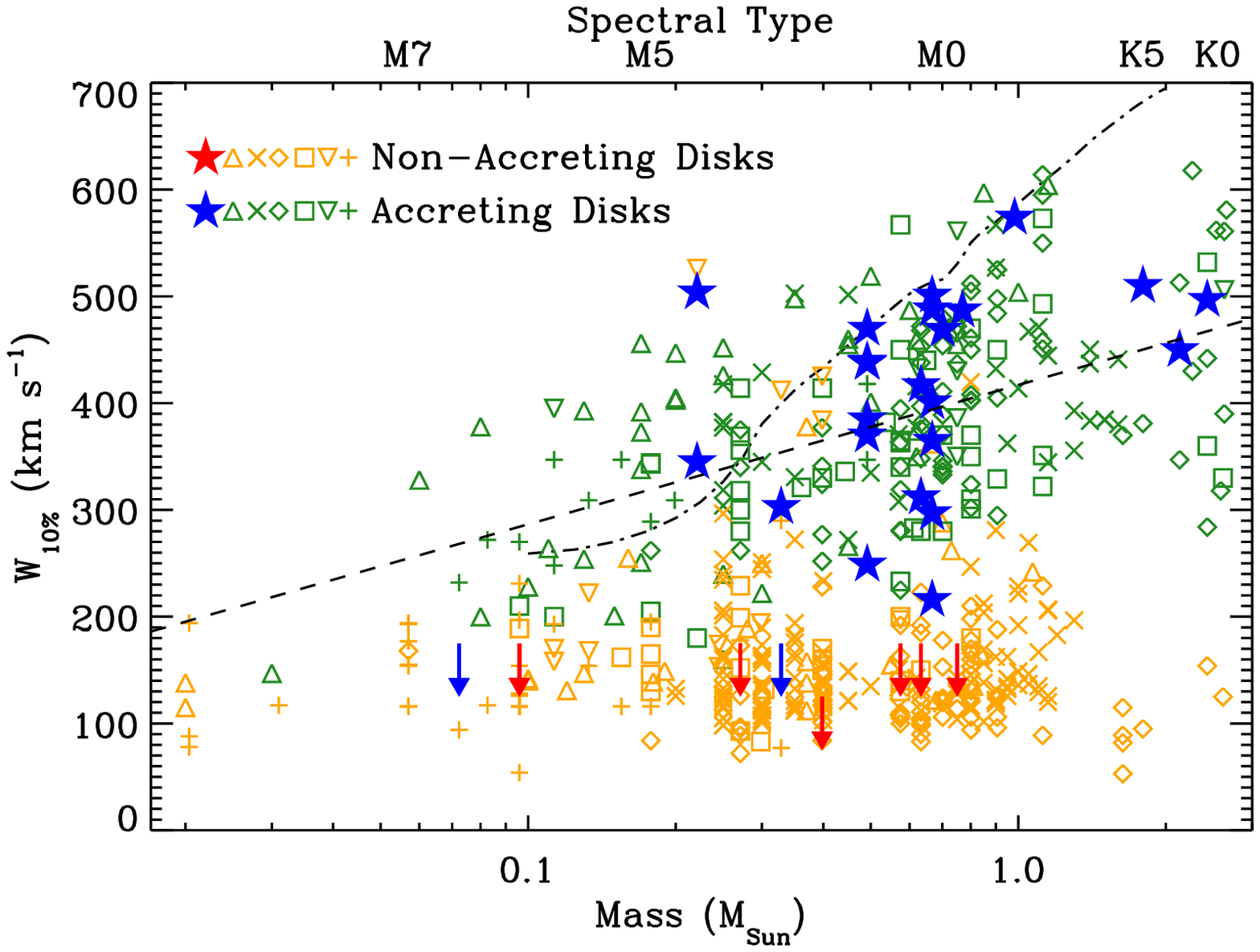}
\caption{{\it Left: } \ha\ EW measurement as a function of stellar mass for T\,Tauri stars. Blue and red stars indicate the weighted average (excluding upper limits) for accreting and non-accreting objects in our sample, respectively. Open symbols represent sample of WTTS, CTTS and transition disks in multiple star-forming regions taken from the literature \citep{cieza10, cieza12, costigan12, nguyen12, manara13, alcala14, frasca15}. Dark green and orange symbols distinguish accreting from non-accreting objects, the same classification taken from these studies. The dashed and dotted curves represent the EW accretion criterion of \citet{barrado03} and 20 times that value, respectively. {\it Right: } \tw\ \ha\ measurements as a function of stellar mass and spectral type for T\,Tauri stars. Symbols are the same as in the left panel. Objects for which only upper limits were obtained in most epochs are shown as upper limits. The dot-dashed curve represents 1.5$v_{ff}$ based on interpolated stellar radii from \citet{siess00} and assuming free-fall from a distance of 5$R_\star$. The dashed line is the result of log-linear fit to all accretors.}\label{fig:dep_w10}
\end{figure*}

The monitoring campaign that we have conducted allows us to revisit in more depth the usefulness of the traditional criteria based on the \ha\ emission line to assess the accretion status of TTS. 

First of all, we notice that the \ha\ EW is over the spectral type-dependent threshold proposed by \citet{barrado03} in 83.5\% (resp. 79.6\%) of all individual spectra of confirmed (resp.) likely accretors. The corresponding fraction of spectra where \tw\ exceeds 200\,\kms\ is 95.0\% (resp. 89.6\%). In other words, roughly 10\% and 20\% of individual observations would lead to an incorrect classification based only on the \tw\ and EW criterion, respectively. In other words, the discrimitative power of traditional \ha-based criteria is strong but imperfect. While the fractions quoted above could be biased due to the make up of our sample, they provide a reasonable order-of-magnitude estimate of the error rate of these criteria. Since the initial reconnaissance of new, large young stellar populations still often relies on low- to moderate-resolution optical spectroscopy for practical reasons, it is important to understand the limits inherent to the traditional \ha\ criteria. In particular, while the EW criterion has a relatively uniform misclassification rate as a function of spectral type, the \tw\ criterion only correctly senses accretion in less than two thirds of all objects with a spectral type later than M3. We revisit below this issue.

To place our results in the broader context, we have assembled a large sample of TTS with measured \ha\ EW and/or \tw\ from the recent literature, specifically using the results from \cite{cieza10, cieza12}, \cite{costigan12}, \cite{nguyen12}, \cite{manara13}, \cite{alcala14}, and \cite{frasca15}. In each of these studies, the accretion status of each object has been determined based on both \tw\ estimates and the presence of other emission lines, essentially following the same methodology as ours. We are interested in probing possible dependencies with stellar mass, but this quantity has not always been estimated in these studies. Since deriving the mass of TTS is a model-dependent proposition based on placing the stars in the Hertzprung-Russell diagram \cite[e.g.,][]{hillenbrand04}, the ideal approach would be to use a uniform method to derive masses to all objects considered here. However, the availability of the necessary photometry and spectral information is far from uniform, leading to uncertainties. Furthermore, the impact of the detailed accretion history, activity, rotation and magnetic field on the location of TTS in the HR diagram may be larger than previously thought \citep{baraffe12, somers14, jackson14, bouvier16}. While we cannot escape these issues, we decided to adopt the stellar masses adopted in the studies referred above whenever available. In all other cases, including for our sample, we used for convenience the relationship between spectral type and stellar mass proposed by \cite{kirk11}, which assume a common system's age of 1\,Myr for all targets. While uncertainties on order 50 per cent are introduced, we are confident that the qualitative trends (or absence thereof) that we find in our analysis are robust. However, we stress that uncertainties on quantitative trends, particularly dependencies on stellar mass, should be taken with caution.

Let us first consider the \ha\ EW, illustrated in the left panel of Figure\,\ref{fig:dep_w10}. While the criterion proposed by \cite{barrado03} marks the approximate transition between accretors and non-accretors, the upper envelope of the non-accretors lies significantly above the lower envelope of the accretors, by up to a factor of $\sim$2. In other words, there is no definitive EW discriminating threshold between accretors and non-accretors, as we already pointed out from our own dataset. As illustrated in our survey, objects with weak, but broad, \ha\ line underline the importance of moderate-to-high resolution spectroscopy. Conversely, intense flares in disk-less but chromospherically active TTS has occasionally led some stars to cross the traditional accretion threshold \citep[e.g.][]{riviere13}. While this has been pointed out in the past \citep[e.g.,][]{fang13}, this analysis confirms that there is a "gray area" of overlap between accretors and non-accretors based on \ha\ EW. This does not mean that previous criteria are poor: only about 2\% of all non-accretors have an EW that exceeds the threshold proposed by \cite{barrado03}, while about 13\% of all accretors fall below it. Therefore, both the purity and completeness of this criterion are high, and it is very likely to yield the correct classification for any one object. However, applying it for a large sample is liable to produce a small number of incorrectly-classified objects, especially bona fide accretors masquerading as non-accretors. As an aside, we note that the EW upper and lower envelopes for accretors follows the same shape, indicating that the range of line EW is independent of stellar mass, hence luminosity. 

Similarly, we plot \tw\ as a function of stellar mass for the same samples from the literature in the right panel of Figure\,\ref{fig:dep_w10}. Focusing first on the non-accretors, we find no significant dependency of their \tw\ with stellar mass, with a median value of $\approx$130\,\kms, respectively. A handful of objects have much broader lines, up to $\gtrsim$400\,\kms, but these are either known or suspected spectroscopic binaries, for which line broadening is induced by the blending of two lines that are shifted as a result of orbital motion \citep[e.g.][]{manara13}. The 90$^{\rm th}$ and 95$^{\rm th}$ percentiles (205 and 245\,\kms, respectively) are in line with the previously proposed upper limits \citep[200-270\,\kms][]{white03, natta04}. However, we emphasize that, once a sufficiently large sample is gathered, no clear upper cut-off in the \tw\ associated with chromospheric activity appears, so that outlier non-accretors can overlap with accretors in \tw. This is especially true for the lower mass objects, with $M \lesssim 0.5\,M_\odot$, whereas a cleaner gap exists for $M \gtrsim 1\,M_\odot$. This is in line with our finding that later type stars often have narrow \ha\ lines while clearly accreting based on other emission lines.

In summary, this analysis confirms that the \ha\ line alone is an imperfect discriminator of accretion in TTS. Even the use of \tw\ measurements, which has taken a larger role in recent years, is insufficient to definitely assess the accretion status of a given object, especially if its mass is $\lesssim 0.5\,M_\odot$. On the other hand, the discriminative power of \tw\ is much higher for $M_\star \gtrsim 1\,M_\odot$. Interpreting occasional outliers from large-scale low-resolution optical surveys, the usual first step in characterizing a new, distant star-forming region, is therefore fraught with ambiguity and should be done with great caution.

%_ _ _ _ _ _ _ _ _ _ _ _ _ _ _ _ _ _ _ _ _ _ _ _ _ _ _ _ _ _ _ _ _ _ _ _ _ _ _ _ _ 

\subsection{\ha\ as a quantitative tracer of the mass accretion rate for TTS}
\label{subsec:discuss_timeevol}

\subsubsection{Ensemble correlations}

Many attempts have been made at converting the \ha\ line properties into an estimate of the accretion rate on the central star. While these efforts have not been very successful for the EW because of other contributions to the line emission \citep[e.g.,][]{antoniucci11}, \cite{natta04} first noted that the line \tw\ is well correlated with the accretion rate, a conclusion further supported by \cite{herczeg08} albeit with a different quantitative relationship. Here we use the results of our survey, both at an ensemble level and through temporal variations for individual objects, to revisit this conclusion.

We first focus on the time-averaged \tw\ measurements for accretors and find that a significant correlation with stellar mass is evident in Figure\,\ref{fig:dep_w10}. For instance, a Spearman rank correlation test returns a false alarm probability, i.e., the probability that the quantities are not correlated, of $p < 6.3\, 10^{-10}$. Assuming a simple log-linear fit (\tw$ = \alpha + \beta \log(M_\star)$), the ensemble of all accretors is characterized by $\beta \approx 125$\,\kms\,dex$^{-1}$, with an intrinsic scatter of $\sim$85\,\kms\ that is independent of stellar mass. To evaluate the influence of our mass determination method, we also repeated the same fit using only stars for which stellar masses were derived from placing the stars in the Hertzprung-Russell diagram in previous studies. This yields a very similar slope of $\beta \approx 130$\,\kms\,dex$^{-1}$, confirming that our simple method to estimate stellar masses does not introduce significant biases in this correlation.

To interpret this correlation, one must keep in mind that correlation does not imply causation. In particular, separate correlations with a "hidden" third quantity could explain the observed correlation between stellar mass and average \tw. A quantity that could naturally play such a role is the stellar accretion rate, which has been found to correlate with both stellar mass and \tw. The stellar accretion rate has been found to follow a $\dot{M}_{acc} \propto M_\star^{2}$ relationship \cite[see][for a recent compilation]{hartmann16}, although there is a range of proposed power law index depending on selected samples and methodologies \citep{muzerolle03, barentsen11, rigliaco11, venuti14}. Similarly, a log-linear relationship has been found between the accretion rate and \ha\ \tw, with $\log{\dot{M}_{acc}} \propto [0.5 .. 2] 10^{-2}$\,\tw\ \citep{natta04, herczeg08, alcala14}. Notwithstanding the caveat that this result is often based on non-simultaneous measurements, combining those two dependencies, one thus expects a correlation between \tw\ and $\log{M_\star}$ characterized by a slope of [100 .. 400]\,\kms\,dex$^{-1}$. While this derivation is fraught with possible underlying biases, the slope that we observe in the \tw--$M_\star$ diagram is thus consistent with previously derived correlations, leaving open the question of the meaningfulness of this correlation.

Generally speaking, if three quantities are each correlated with one another, it is most likely that one of the correlations is merely a consequence of the other two, rather than indicative of a physical causality. Here, the quantities under consideration are stellar mass, stellar accretion rate and \ha\ \tw. Arguably the most intuitive physical connection one might expect is between stellar mass and \tw. In a schematic picture where mass accretion onto the central star occurs near the free-fall velocity, one expects lower-mass stars to have smaller accretion flow velocities. To explore this possibility in a quantitative manner, we compute 1.5$v_{ff}$ as a reasonable proxy for the resulting line \tw. Indeed, only the very extreme tails of the line profile can correspond to velocities of $\pm v_{ff}$ and projection effects further reduce the resulting radial velocity offsets as seen from the observer's point of view. We further assume that the free fall initiates at a distance from the star of 5\,$R_\star$ following previous studies \citep{gullbring98} and adopt the stellar radii from the evolutionary models of \cite{siess00}. The resulting stellar mass--\tw\ relationship is markedly steeper ($\beta \approx 400$\,\kms\,dex$^{-1}$, see Figure\,\ref{fig:dep_w10}) than the correlation we observe, suggesting that the latter stems from a different mechanism. This leads us to conclude the stellar mass--\tw\ correlation is a good candidate for being a mere consequence of other, physically grounded, correlations. 

In turn, this analysis supports the idea that the stellar accretion rate is correlated with both stellar mass and \tw. While the debate about the physical interpretation of the correlation between the first two quantities is still ongoing, several possibilities have been proposed \citep[e.g.,][]{alexander06, gregory06, vorobyov08, manara16}. On the other hand, the physical explanation of the correlation between accretion rate and \tw\ remains elusive to the point that some authors argue that there is no such mechanism \cite{hartmann16}. Furthermore, the fact that this correlation is generally based on (often non-simultaneous) single measurements of ensemble of stars invites caution in its use to quantitatively estimate accretion rate in TTS, especially in the face of the large scatter in the observed correlation \citep{fang13, alcala14}. Nonetheless, because of the relative ease with which \tw\ can be measured, it is often used as a robust quantitative proxy for the mass accretion rate. However, the validity of this approach remains to be demonstrated.

\subsubsection{Temporal variability of individual objects}

\begin{figure}
\includegraphics[width=0.49\textwidth]{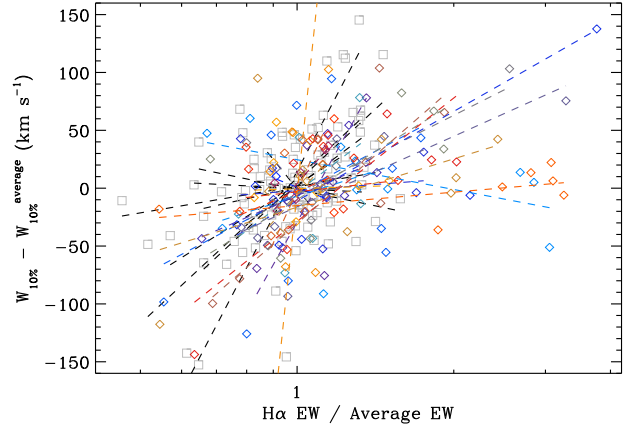}
\caption{EW and \tw\ \ha\ measurement for all objects in our sample (colored diamonds) with at least two epochs where \tw\ could be measured for all accretors in our sample; each color corresponds to a different target. Both quantities are normalized by the weighted average for each object to focus on relative changes in the line strength and overall width. For each target, the result of log-linear fits is also shown as a dashed line spanning the same range of (normalized) EW for the target: objects with the largest excursions in EW are associated to the longest lines. Gray squares represent individual measurement by accretors studied in \citet{costigan12}; black dashed lines are similar log-linear fits for these targets.}\label{fig:ew_tw}
\end{figure}

We can take advantage of our multi-epoch monitoring campaign to shed new light on the accretion rate--\tw\ correlation by considering one star at a time. In doing so, we hold fixed a number of potentially important parameters, such as evolutionary state, structure of the inner disk and viewing geometry, whose large variety is likely responsible for a significant fraction of the scatter of ensemble correlations\footnote{Temporal variability of the accretion rate can only account for a fraction of the total scatter in such correlations \citep{venuti14}.}. We thus focus on the temporal variability of the line measurements for each of the accreting TTS in our survey. This subsample contains 23 objects, with a median of 7 observations per target. Although the relationship between line profile and accretion rate is likely complex, comparing the various observations of the same object alleviates many of the uncertainties listed above. Specifically, if \tw\ is indeed a good quantitative tracer of the accretion rate, fluctuations in \tw\ would readily reveal fluctuations in $\dot{M}_{acc}$. The latter is proportional to $L_{acc}$, itself correlated with the luminosity of accretion-driven emission \citep[but see][]{mendigutia15} and thus we would expect \tw\ to be correlated with $L_{H\alpha}$. While we have not measured that quantity, we note that the continuum brightness variability due to changes in veiling for most TTS are generally $\leq$50 per cent for non-outbursting TTS \citep[e.g.,][]{stauffer16}, smaller than the amplitude of the line EW fluctuations for most of the accretors in our sample. Therefore, our EW measurements are a reasonable proxy for the line luminosity. We may thus expect a positive correlation between EW and \tw, which would be partially attenuated if accretion variability induces significant, correlated veiling changes \cite{alencar02}.

Plotting every epoch for every target in an EW--\tw\ diagram reveals a scatter plot with no obvious correlation. This is consistent with the lack of obvious correlation between these quantities found by \cite{costigan12, costigan14}, for instance. However, since we are primarily interested in relative changes in the line strength and widths, we proceed to subtract (divide by) the weighted mean of all \tw\ (EW) from each individual measurements. The results are shown in Figure\,\ref{fig:ew_tw}, in which we also plot the individual measurements from \cite{costigan12}. Despite the large scatter, we find a strong correlation between the normalized EW and \tw\ measurements, with a probability of being uncorrelated of less than 5--6\,10$^{-6}$ for both our dataset and that of \cite{costigan12}. Combining the two samples, the null hypothesis that the two quantities are uncorrelated has a probability of $\approx 10^{-10}$. In other words, for the vast majority of CTTS, when the \ha\ line becomes stronger, it gets broader: the line profile changes in a correlated manner with the line intensity.

The limited sample number of observations per star, coupled with the limited amplitude in both line strength and width, is such that the significance of a putative correlation (as measured by a Spearman test, for instance) is modest at best for each individual target. Nonetheless, given the strong correlation over the entire sample, we proceed to perform least-squares log-linear fitting to each target to estimate the slope of the \tw--log(EW) correlation, $\gamma$. The resulting trend is systematically in the same direction, and of similar slope, for all targets. The weighted average for the correlation slopes over all accretors in our sample is $\overline{\gamma} = 145\pm10$\,\kms\,dex$^{-1}$, where the uncertainty is the standard deviation of the mean. The scatter about the mean is $\approx 140$\,\kms\ for objects whose EW varies by at least a factor of two within our campaign. Similarly, the sample of accretors monitored by \cite{costigan12} in the Chamaeleon star-forming region has a similar mean value ($\gamma_{\mathrm Cha} = 145\pm100$\,\kms\,dex$^{-1}$, albeit with a larger intrinsic scatter of 300\,\kms). The observed star-to-star scatter is a factor of $\approx3$ higher than is expected given the formal uncertainties on the log-linear fits, indicating that there is an intrinsic diversity in the slope of this correlation, as well as more complex variability patterns. Nonetheless, all but two targets in our sample have a positive correlation. Comforted by the similarity between the Taurus and Chamaelon samples, we conclude that the correlation between normalized EW and \tw\ applies to accreting TTS at large and, thus, reveals an underlying physical process.

We have found that the \ha\ line is characterized by a larger \tw\ when it is stronger. Since this is not true for simple amplitude fluctuations of a line following a Gaussian profile, the shape of the line profile must be somehow correlated with the strength of the line itself. One way the line profile can change is through changes in the level of self-absorption of the line, whereby the line appears stronger when absorption is reduced. However, if such changes occur without modification in the high-velocity wings, they would result in a lower \tw\ since the line peak is then higher. Therefore, increased line strength is correlated with enhanced emission at high velocities. Enhanced mass loading in the accretion funnels, resulting in more material traveling near the free-fall velocity, is one mechanism to achieve this, although we stress that the line radiative transfer problem in this environment is sufficiently complex that qualitative arguments may turn out to be overly simplistic. Unfortunately, our moderate spectral resolution does not allow us to characterize further the exact nature of the line profile change. A high-resolution follow-up study would be necessary to probe this aspect. Nonetheless, the discovery of the systematic correlation of the \ha\ EW with its \tw\ is strongly suggestive of the fact that changes in the accretion rate are directly responsible for changes in \tw.

%________________________________________________________________

\section{Conclusion}
\label{sec:conclu}

We have conducted a 12-month monitoring campaign of a sample of 33 TTS in the Taurus star-forming region, all but two of which host a circumstellar disk. These disks span the range of protoplanetary disks, including full-fledged disks, transition disks with inner holes, and "anemic" disks with only modest amount of circumstellar material. The sample includes 9 objects whose previously measured \ha\ EW suggest that they are non-accreting despite having circumstellar dust, thus hinting at the possibility that their disk are passive. We therefore set out to assess the accretion status of these objects. An additional goal of our program is to identify objects in which accretion flickers on and off on timescales of days to 1\,yr.

We have performed medium-resolution ($R\sim2000$-5000) optical spectroscopy to measured the EW and \tw\ of the \ha\ line. We use these measurements in combination with the presence of other emission lines to determine whether each target was accreting at any epoch. Since our methodology relies on lower resolution spectroscopy than is typical in this context, we have used a grid of simulations to demonstrate our ability to retrieve the intrinsic \tw\ of even weak lines from spectra at that resolution. Through this analysis, we have quantified the bias in \tw\ that is mainly introduced by weak absorption features in the underlying photospheric spectrum.

We established the accretion status of our target based on an array of tracers included in our spectra but not the \ha\ line in order to assess the reliability of the latter. We find that the \ha\ line is stronger (wider) among accreting TTS than the minimum thresholds that have been proposed in the past in 80\% (90\%) of cases in our study, emphasizing that the criteria are good but imperfect. As a result, accretion status derived from low-resolution \ha\ spectroscopy are therefore fraught with a non-negligible fraction of errors, which can have significant consequences when studying new, distant young stellar populations.

Besides the two disk-less targets, we found 6 objects that show no sign of accretion at any epoch during our survey. Two of these are associated with transition disks, and two more with "anemic" disks. In the first case, there is no circumstellar material within the inner few AU, whereas in the second one, the disk is better characterized as a debris disk, explaining the lack of accretion. The remaining two non-accreting disk-bearing systems are binary or high-order multiple systems in which it is probable that the star found to be not accreting is not the one having circumstellar material. In summary, none of these systems is a full-fledged, non-accreting protoplanetary disk. Furthermore, despite marked line variability for all accreting TTS, we find no convincing evidence for flickering accretion, allowing us to place a 2.2 percent upper limit on the duty cycle of accretions "gaps" in TTS if we assume that all TTS undergo such episodes.

Combining the average \ha\ measurements for all objects in our sample with literature data, we further study the imperfect nature of the \ha-based criteria to distinguish accreting from non-accreting TTS. In particular, while the \tw\ measurements of CTTS and WTTS are well separated for objects whose mass is $\gtrsim1\,M_\odot$, the two samples overlap substantially in the low-mass regime. For CTTS, we find a significant correlation between \tw\ and stellar mass which does not seem to be related to the variation in free-fall velocity, suggesting that both quantities are instead independently correlated with a third quantity. The stellar mass accretion rate is a plausible contender for playing such a role.

We take advantage of the time series of our \ha\ measurements for accreting objects to quantitatively study the variability of the line. As in previous studies, we find that most of the variability is established on timescales as short as 3\,d. We show that, for most accretors, the \tw\ of the \ha\ line is positively correlated with its EW, with a slope that is consistent from star to star. We also find similar average and dispersion of these slopes in a sample of CTTS in Chamaeleon that have been similarly monitored. We infer from this pervasive correlation that when variations of the \ha\ line leads to a stronger total luminosity, it is associated with enhanced emission in the high-velocity wings. Although we cannot conclusively prove it with our data, we argue that this is strong evidence in support of the hypothesis that the mass accretion rate on the central star is directly correlated with the \ha\ \tw\ and, thus, that the former can be used to quantitatively estimate the latter. The physical mechanism through which this correlation is established remains to be explained. However, the similarity in the correlation slope over two samples of CTTS suggests a common mechanism, even though line profiles are highly object-dependent.

\section*{Acknowledgements}

We thank the referee for a prompt report that allowed us to strengthen our analysis. We are grateful to the support staff at all observatories where data presented in this study were obtained, and extend particular thanks to the Calar Alto Observatory staff who conducted our observations in service mode and to the daytime and night support staff at the OAN-SPM for facilitating and helping obtain our observations. This article is based on observations made in the Observatorios de Canarias del IAC with the William Herschel Telescope operated on the island of La Palma by the Isaac Newton Group of Telescopes in the Observatorio del Roque de los Muchachos. Based on observations collected at the Centro Astron\'omico Hispano Alem\'an (CAHA) at Calar Alto, operated jointly by the Max-Planck Institut f\"ur Astronomie and the Instituto de Astrof\'{\i}sica de Andaluc\'{\i}a (CSIC). We thank Calar Alto Observatory for allocation of director's discretionary time to this programme. Based upon observations carried out at the Observatorio Astron\'omico Nacional on the Sierra San Pedro M\'artir (OAN-SPM), Baja California, M\'exico. This work was partially supported by funds from the UC Berkeley Undergraduate research Apprenticeship Program and by the Programa de Acceso a Grandes Instalaciones Cient\'{\i}ficas (No. 2009/00157 and 2009/00203), administered by the Instituto de Astrof\'{\i}sica de Canarias and El Ministerio de Educaci\'on y Ciencia, Espa\~na. RJDR was funded through a studentship from the Science and Technology Facilities Council (STFC) ST/F 007124/1.

%%%%%%%%%%%%%%%%%%%%%%%%%%%%%%%%%%%%%%%%%%%%%%%%%%



%\bibliographystyle{mnras}

\begin{thebibliography}{99}
\bibitem[Alcal{\'a} et al.(2017)]{alcala16} Alcal\'a, J. N., Manara, C. F., Natta, A., Frasca, A., Testi, L., Nisini, B., Stelzer, B. et al. (2017) \aap, in press (arxiv:1612.07054)
\bibitem[Alcal{\'a} et al.(2014)]{alcala14} Alcal\'a, J. N., Natta, A., Manara, C. F., Stelzer, B., Frasca, A., Biazzo, K., Covino, E. et al. (2014) \aap, 561, 2
\bibitem[Alencar \& Batalha(2002)]{alencar02}Alencar, S. H. P., \& Batalha, C. (2002) \apj, 571, 378
\bibitem[Alexander et al.(2014)]{alexander14} Alexander, R., Pascucci, I., Andrews, S., Armitage, P., \& Cieza, L. (2014) in {\it Protostars \& Planets VI}, H. Beuther, R. S. Klessen, C. P. Dullemond \& T. Henning eds., Univ. of Arizona Press, Tucson, 475
\bibitem[Allard et al.(1997)]{allard97} Allard, F., Hauschildt, P. H., Alexander, D. R. \& Starrfield, S. (1997) \araa, 35, 137
\bibitem[Alexander \& Artmitage(2006)]{alexander06} Alexander, R. D. \& Armitage, P. J. (2006) \apjl, 639, L83
\bibitem[Andrews \& Williams(2005)]{andrews05} Andrews, S. M., \& Williams, J. P. (2005) \apj, 631, 1134
\bibitem[Antoniucci et al.(2011)]{antoniucci11} Antoniucci, S., Garc\'{\i}a L\'opez, R., Nisini, B., Giannini, T., Lorenzetti, D., Eisl\"offel, J., et al. (2011) \aap, 534, 32
\bibitem[Audard et al.(2014)]{audard14} Audard, M., \'Abrah\'am, P., Dunham, M. M., Green, J. D., Grosso, N., Hamaguchi, K., et al. (2014) in {\it Protostars \& Planets VI} H. Beuther, R. S. Klessen, C. P. Dullemond \& T. Henning eds., Univ. of Arizona Press, Tucson, 387
\bibitem[Baraffe et al.(2012)]{baraffe12} Baraffe, I., Vorobyov, E. \& Chabrier, G. (2012) \apj, 756, 118
\bibitem[barentsen et al.(2011)]{barentsen11} Barentsen, G., Vink, J. S., Drew, J. E., Greimel, R., Wright, N. J., Drake, J. J., et al. (2011) \mnras, 415, 103
\bibitem[Barrado y Navascu\'es \& Mart\'{\i}n(2003)]{barrado03} Barrado y Navascu\'es, D., \& Mart\'{\i}n, E. L. (2003) \aj, 126, 2997
\bibitem[Beckwith et al.(1990)]{beckwith90} Beckwith, S. V. W., Sargent, A. I., Chini, R. S., \& Guesten, R. (1990) \aj, 99, 924
\bibitem[Bertout(1989)]{bertout89} Bertout, C. (1989) \araa, 27, 351
\bibitem[Bouvier et al.(2003)]{bouvier03} Bouvier, J., Grank, K. N., Alencar, S. H. P., Dougados, C., Fern\'andez, M., Basri, G., et al. (2003) \aap, 409, 169
\bibitem[Bouvier et al.(2007)]{bouvier07} Bouvier, J., Alencar, S. H. P., Harries, T. J., Johns-Krull, C. M., \& Romanova, M. M. (2007) in {\it Protostars \& Planets V}, B. Reipurth, D. Jewitt \& K. Keil eds, Univ. of Arizona Press, Tucson, 479
\bibitem[Bouvier et al.(2007)]{bouvier08} Bouvier, J., Alencar, S. H. P., Boutelier, T., Dougados, C., Balog, Z., Grankin, K., et al. (2007) \aap, 463, 1017
\bibitem[Bouvier et al.(2016)]{bouvier16} Bouvier, J., Lanzafame, A. C., Venuti, L., Klutsch, A., Jeffries, R., Frasca, A., Moraux, E. et al. (2016) \aap, 590A, 78
\bibitem[Brice\~no et al.(1998)]{briceno98} Brice\~no, C., Hartmann, L., Stauffer, J., \& Mart\'{\i}n, E. (1998) \aj, 115, 2074
\bibitem[Chelli et al.(1988)]{chelli88} Chelli, A., Zinnecker, H., Carrasco, L., Cruz-Gonzalez, I. \& Perrier, C. (1988) \aap, 207, 46
\bibitem[Cieza et al.(2010)]{cieza10} Cieza, L. A., Schreiber, M. R., Romero, G. A., Mora, M. D., Merin, B., Swift, J. S., Orellana, M. et al. (2010) \apj, 712, 925
\bibitem[Cieza et al.(2012)]{cieza12} Cieza, L.A., Schreiber, M. R., Romero, G. A., Williams, J. P., Rebassa-Mansergas, A. \& Merin, B. (2012) \apj, 750, 157
\bibitem[Cieza et al.(2013)]{cieza13} Cieza, L. A., Olofsson, J., Harvey, P. M., Evans, N. J. I., Najita, J., Henning, T., et al. (2013) \apj, 762, 100
\bibitem[Costigan et al.(2012)]{costigan12} Costigan, G., Scholz, A., Stelzer, B., Ray, T., Vink, J. S., \& Mohanty, S. (2012) \mnras, 427, 1344
\bibitem[Costigan et al.(2014)]{costigan14} Costigan, G., Vink, J. S., Scholz, A., Ray, T., \& Testi, L. (2014) \mnras, 440, 3444
\bibitem[Donati et al.(2013)]{donati13} Donati, J.-F., Gregory, S., Alencar, S., Hussain, G., Bouvier, J., Jardine, M. et al. (2013) \mnras, 436, 881
\bibitem[Espaillat et al.(2014)]{espaillat14} Espaillat, C., Muzerolle, J., Najita, J., Andrews, S., Zhu, Z., Calvet, N., et al. (2014) in {\it Protostars \& Planets VI}, H. Beuther, R. S. Klessen, C. P. Dullemond \& T. Henning eds., Univ. of Arizona Press, Tucson, 497
\bibitem[Fang et al.(2013)]{fang13} Fang, M., Kim, J. S., van Boekel, R., Sicilia-Aguilar, A., Henning, T. \& Flaherty, K. (2013) \apjs, 207, 5
\bibitem[Fang et al.(2009)]{fang09} Fang, M., van Boekel, R., Wang, W., Carmona, A., Sicilia-Aguilar, A. \& Henning, Th. (2009) \aap, 504, 461
\bibitem[Frasca et al.(2015)]{frasca15} Frasca, A., Biazzo, K., Lanzafame, A. C., Alcal\'a, J. M., Brugaletta, E., Klutsch, A., Stelzer, B. et al. (2015) \aap, 575, 4
\bibitem[Frasca et al.(2017)]{frasca17} Frasca, A., Biazzo, K., Alcal\'a, J. M., Manara, C. F., Stelzer, B., Covino, E. \& Antoniucci, S. (2017) \aap, in press (arxiv:1703.01251)
\bibitem[Furlan et al.(2009)]{furlan09} Furlan, E., Forrest, W. J., Sargent, B. A., Manoj, P., Kim, K. H. \& Watson, D. M. (2009) \apj, 706. 1194
\bibitem[Gregory et al.(2006)]{gregory06} Gregory, S. G., Jardine, M., Simpson, I. \& Donati, J.-F. (2006) \mnras, 371, 999
\bibitem[Gullbring et al.(1998)]{gullbring98} Gullbring, E., Hartmann, L., Brice\~no, C., \& Calvet, N. (1998) \apj, 492, 323
\bibitem[Haisch et al.(2001)]{haisch01} Haisch, K. E., Lada, E. A., \& Lada, C. J. (2001) \apjl, 553, L153
\bibitem[Hardy et al.(2015)]{hardy15} Hardy, A., Caceres. C., Schreiber, M. R., Cieza, L., Alexander, R. D., Canovas, H., Williams, J. P., et al. (2015) \aap, 583, 66
\bibitem[Hartigan et al.(1995)]{hartigan95} Hartigan, P., Edwards, S. \& Ghandour, L. (1995) \apj, 452, 736
\bibitem[Hartigan et al.(1994)]{hartigan94} Hartigan, P., Strom, K. M., \& Strom, S. E. (1994) \apj, 427, 961
\bibitem[Hartmann et al.(2016)]{hartmann16} Hartmann, L., Hercez, G. \& Calvet, N. (2016) \mnras, 54, 135
\bibitem[Herbig(1960)]{herbig60} Herbig, G. (1960) \apjs, 4, 337
\bibitem[Herbig \& Bell(1988)]{hbc88} Herbig, G. H., \& Bell, K. R. (1988) in {\it Lick Observatory Bulletin}, 1111, 90
\bibitem[Herczeg \& Hillenbrand(2008)]{herczeg08} Herczeg, G. J., \& Hillenbrand, L. A. (2008) \apj, 681, 594
\bibitem[Herczeg \& Hillenbrand(2014)]{herczeg14} Herczeg, G. J., \& Hillenbrand, L. A. (2014) \apj, 786, 97
\bibitem[Hern\'andez et al.(2014)]{hernandez14} Hern\'andez, J., Calvet, N., Perez, A., Brice\~no, C., Olguin, L., Contreras, M. et al. (2014) \apjl, 630, L185
\bibitem[Hillenbrand \& White(2004)]{hillenbrand04} Hillenbrand, L. A. \& White, R. J. (2004) \apj, 604, 741
\bibitem[Howard et al.(2013)]{howard13} Howard, C. D., Sandell, G., Vacca, W. D., Duchene, G., Mathews, G., Augereau, J.-C., et al. (2013) \apj, 776(1), 21
\bibitem[Ingleby et al.(2013)]{ingleby13} Ingleby, L., Calvet, N., Herczeg, G., Blaty, A., Walter, F., Ardila, D., et al. (2013) \apj, 767, 112
\bibitem[Jackson \& Jeffries(2014)]{jackson14} Jackson, R. J. \& Jeffries, R. D. (2014) \mnras, 445, 4306
\bibitem[Jayawardhana et al.(2006)]{jayawardhana06} Jayawardhana, R., Coffey, J., Scholz, A., Brandeker, A. \& van Kerkwijk, M. H. (2006) \apj, 648, 1206
\bibitem[Joy(1945)]{joy45} Joy, A. (1945) \pasp, 57, 171
\bibitem[Kirk \& Myers(2011)]{kirk11} Kirk, H. \& Myers, P. C. (2011) \apj, 727, 64
\bibitem[Koresko et al.(1997)]{koresko97} Koresko, C. D., Herbst, T. M. \& Leinert, Ch. (1997) \apj, 480, 741
\bibitem[Kraus et al.(2015)]{kraus15} Kraus, A. L., Andrews, S. M., Bowler, B. P., Herczeg, G., Ireland, M. J., Liu, M. C., et al. (2015) \apjl, 798(1), L23
\bibitem[Luhman et al.(2006)]{luhman06} Luhman, K. L., Whitney, B. A., Meade, M. R., Babler, B. L., Indebetouw, R., Bracker, S., \& Churchwell, E. B. (2006) \apj, 647, 1180
\bibitem[Luhman et al.(2010)]{luhman10} Luhman, K. L., Allen, P. R., Espaillat, C., Hartmann, L., \& Calvet, N. (2010) \apjs, 186, 111
\bibitem[Manara et al.(2016)]{manara16} Manara, C. F., Rosotti, G., Testi, L., Natta, A., Alcal\'a, J. M., Williams, J. P., et al. (2016) \aap L, 591, L3
\bibitem[Manara et al.(2013)]{manara13} Manara, C. F., Testi, L., Rigliaco, E., Alcal\'a, J. M., Natta, A., Stelzer, B., Biazzo, K. et al. (2013) \aap, 551, 107
\bibitem[Mathews et al.(2013)]{mathews13} Mathews, G. S., Pinte, C., Duch\^ene, G., Williams, J. P. \& M\'enard, F. (2013) \aap, 558, 66
\bibitem[McCabe et al.(2006)]{mccabe06} McCabe, C., Ghez, A. M., Prato, L., Duch\^ene, G., Fisher, R. S., \& Telesco, C. (2006) \apj, 636, 932
\bibitem[Mendigut\'{\i}a et al.(2015)]{mendigutia15} Mendigut\'{\i}a, I., Oudmaijer, R. D., Rigliaco, E., Fairlamb, J. R., Calvet, N., Muzerolle, J., et al. (2015) \mnras, 452, 2837
\bibitem[Mundt et al.(1983)]{mundt83} Mundt, R., Walter F., Feigelson, E., Finkenzeller, U., Herbig, G., Odell, A. (1983) \apj, 269, 229
\bibitem[Murphy et al.(2011)]{murphy11} Murphy, S. J., Lawson, W. A., Bessell, M. S., \& Bayliss, D. D. R. (2011) \mnras L, 411(1), L51
\bibitem[Muzerolle et al.(2003)]{muzerolle03} Muzerolle, J., Hillenbrand, L., Calvet, N., Brice\~{n}o, C., Hartmann, L. (2003) \apj, 592, 266
\bibitem[Muzerolle et al.(1998)]{muzerolle98} Muzerolle, J., Hartmann, L. \& Calvet, N. (1998) \aj, 116, 2965
\bibitem[Najita et al.(2007)]{najita07} Najita, J. R., Strom, S. E., \& Muzerolle, J. (2007) \mnras, 378, 369
\bibitem[Natta et al.(2004)]{natta04} Natta, A., Testi, L., Muzerolle, J., Randich, S., Comer\'on, F., \& Persi, P. (2004) \aap, 424, 603
\bibitem[Nguyen et al.(2009)]{nguyen09} Nguyen, D. C., Scholz, A., van Kerkwijk, M. H., Jayawardhana, R., \& Brandeker, A. (2009) \apjl, 694, L153
\bibitem[Nguyen et al.(2012)]{nguyen12} Nguyen, D. C., Brandeker, A., van Kerkwijk, M. H., \& Jayawardhana, R. (2012) \apj, 745, 119
\bibitem[Padgett et al.(2006)]{padgett06} Padgett, D. L., Cieza, L., Stapelfeldt, K. R., Evans, N. J., Koerner, D., Sargent, A., et al. (2006) \apj, 645, 1283
\bibitem[Petterson et al.(2014)]{petterson14} Pettersson, B., Armond, T. \& Reipurth, B. (2014) \aap, 570, A30
\bibitem[Ribas et al.(2014)]{ribas14} Ribas, \'A., Mer\'{\i}n, B., Bouy, H., \& Maud, L. T. (2014) \aap, 561, 54
\bibitem[Rigliaco et al.(2012)]{rigliaco12} Rigliaco, E., Natta, A., Testi, L., Randich, S., Alacal\`a, J. M., Covino, E. \& Stelzer, B. (2012) \aap, 548, A56
\bibitem[Rigliaco et al.(2011)]{rigliaco11} Rigliaco, E., Natta, A., Randich, S., Testi, L. \& Biazzo, K. (2011) \aap, 524A, 47
\bibitem[Riviere-Marichalar et al.(2015)]{riviere15} Riviere-Marichalar, P., Elliott, P., Rebollido, I., Bayo, A., Ribas, A., Mer\'{\i}n, B., et al. (2015) \aap, 584, A22
\bibitem[S{\'a}nchez-Bl{\'a}quez et al.(2006)]{sanchez06} S{\'a}nchez-Bl{\'a}quez, P., Peletier, R. F., Jim{\'e}nez-Vicente, J., Cardiel, N., Cenarro, A. J., Falc{\'o}n-Barroso, J., et al. (2006), \mnras, 371, 703
\bibitem[Riviere-Marichalar et al.(2013)]{riviere13} Riviere-Marichalar, P., Pinte, C., Barrado, D., Thi, W. F., Eiroa, C., Kamp, I., Montesinos, B. et al. (2013) \aap, 555, 67
\bibitem[Siess et al.(2000)]{siess00} Siess, L., Dufour, E. \& Forestini, M. (2000) \aap, 358, 593
\bibitem[Somers \& Pinsonneault(2014)]{somers14} Somers, G. \& Pinsonneault (2014) \apj, 790, 72
\bibitem[Stauffer et al.(2016)]{stauffer16} Stauffer, J., Cody, A. M., Rebull, L., Hillenbrand, L. A., Turner, N. J., Carpenter, J., Carey, S. et al. (2016) \aj, 151, 60
\bibitem[Strom \& Strom(1994)]{strom94} Strom, K. M., \& Strom, S. E. (1994) \apj, 424, 237
\bibitem[Valenti et al.(1993)]{valenti93} Valenti, J. A., Basri, G. \& Johns, C. M. (1993) \aj, 106, 2024
\bibitem[Valenti et al.(2003)]{valenti03} Valenti, J. A., Fallon, A. A. \& Johns-Krull, C. M. (2003) \apjs, 147, 305
\bibitem[Venuti et al.(2014)]{venuti14} Venuti, L., Bouvier, J., Flaccomio, E., Alencar, S. H. P., Irwin, J., Stauffer, J. R., et al. (2014) \aap, 570, A82
\bibitem[Vorobyov \& Basu(2008)]{vorobyov08} Vorobyov, E. I. \& Basu, S. (2008) \apjl, 676, L139
\bibitem[Wahhaj et al.(2010)]{wahhaj10} Wahhaj, Z., Cieza, L., Koerner, D. W., Stapelfeldt, K. R., Padgett, D. L., Case, A., et al. (2010) \apj, 724, 835
\bibitem[White \& Ghez(2001)]{white01} White, R. J., \& Ghez, A. M. (2001) \apj, 556, 265
\bibitem[White \& Basri(2003)]{white03} White, R. J., \& Basri, G. (2003) \apj, 582, 1109
\bibitem[Wichmann et al.(1996)]{wichmann96} Wichmann, R., Krautter, J., Schmitt, J. H. M. M., Neuhaeuser, R., Alcal\'a, J. M., Zinnecker, H., et al. (1996) \aap, 312, 439
\bibitem[Williams \& Cieza(2011)]{williams11} Williams, J. P., \& Cieza, L. A. (2011) \araa, 49, 67
\end{thebibliography}


%%%%%%%%%%%%%%%%%%%%%%%%%%%%%%%%%%%%%%%%%%%%%%%%%%

%%%%%%%%%%%%%%%%% APPENDICES %%%%%%%%%%%%%%%%%%%%%

%\appendix


%If you want to present additional material which would interrupt the flow of the main paper, it can be placed in an Appendix which appears after the list of references.

%%%%%%%%%%%%%%%%%%%%%%%%%%%%%%%%%%%%%%%%%%%%%%%%%%

%\begin{table}
%\caption{Emission line measurements}\label{tab:res}
%\begin{tabular}{lcccl}
%\hline
%Target & Date & EW(\ha) & \tw\ & Other lines \\
% &  & [\AA] & [\kms] & \\
%\hline
%\input{table_new.tex}
%\hline
%\end{tabular}
%%\tablecomments{The last line for each object provides the average and standard deviation of all EW  and \tw\ measurements (or the median upper limit on \tw\ in the case of objects where most spectra did not yield a detection).}
%\end{table}

% Don't change these lines
\bsp	% typesetting comment
\label{lastpage}
\end{document}